\documentstyle[preprint,aps,psfig]{revtex}
\begin{document}
\preprint{\vbox{ \hfill MKPH-T-96-26 }}
\title{Electromagnetic Production of the Hypertriton}
\author{T.\ Mart$^{a,b}$, L. Tiator$^{b}$, D. Drechsel$^{b}$, and 
        C.\ Bennhold$^{c}$}
\address{$^{a}$Jurusan Fisika, FMIPA, Universitas Indonesia, 
         Depok 16424, Indonesia\\
         $^{b}$Institut f\"ur Kernphysik, Johannes Gutenberg-Universit\"at, 
         55099 Mainz, Germany\\ 
         $^{c}$Center of Nuclear Studies, Department of Physics, 
         The George Washington University, 
         Washington, D. C. 20052}
\date{\today}
\maketitle
\begin{abstract}
Kaon photoproduction on $^3$He,
$\gamma ~+~ ^3{\mathrm He} \longrightarrow K^{+} +~ ^3_{\Lambda}{\mathrm H}$, 
is studied in the framework of the impulse approximation.
Realistic $^3$He wave functions obtained as solutions of
Faddeev equations with the Reid soft-core potential are used along 
with different $^3_{\Lambda}$H wave functions. 
Results are compared for several elementary operator models, 
which can successfully describe the elementary kaon production off the proton 
up to a photon lab energy of $k = 2.2 $ GeV. It is found that the
corresponding cross sections are small, of the order of several nanobarns.
It is also shown that the influence of Fermi motion is important,  
while the effect of different off-shell assumptions on the
cross section is not too significant.\\
PACS number(s) : {13.60.Le, 25.20.Lj, 21.80.+a}

\end{abstract}

\section{INTRODUCTION}
With the start of experimental activities at Jefferson Lab, the 
electromagnetic production of hypernuclei will become experimentally 
feasible. This reaction offers a particularly efficient tool to study 
the production and interactions of hyperons in the nuclear medium. 
The reaction is of special interest in the case of the lightest 
hypernucleus, the hypertriton $^3_{\Lambda}$H. Studies of the 
hypertriton can provide relevant new information on the $YN$ 
interaction, which up to now is only poorly known from the available 
$YN$ scattering data. Furthermore, with the hypertriton being the lightest
hypernucleus, it is obviously the first system in which the $YN$ 
potential, including the interesting $\Lambda$-$\Sigma$ conversion potential, 
can be tested in the nuclear environment. This is also supported by the fact
that neither the $\Lambda N$ nor the $\Sigma N$ interactions are 
sufficiently strong to produce a bound two-body system. Therefore 
the hypertriton will play an important role in hypernuclear physics,
similar to the deuteron in nuclear physics.

Recently, the Bochum Group \cite{miyagawa} has investigated  the 
hypertriton using the J\"ulich hyperon-nucleon potential in the 
one-boson-exchange (OBE) parametrization (model $\tilde{\mathrm A}$ of 
Ref. \cite{speth}) combined with various realistic $NN$ interactions. 
They found that with this potential the hypertriton turns out to be 
unbound. Only an increase by about 4\% in the J\"ulich potential 
(multiplication of the $^1S_0~YN$ partial wave by a factor of 1.04) 
leads to a bound state for the hypertriton. On the other hand, 
the use of the Nijmegen hyperon-nucleon potential \cite{nijmegen1} in the
same calculation \cite{miyagawa2} leads to a bound hypertriton. 
Clearly, significant improvement is still needed in the hyperon-nucleon 
force sector, where in contrast to the nucleon-nucleon sector
the dominant one-pion-exchange (OPE) tensor force is not present
since the lambda ($I=0$) and the nucleon ($I=\frac{1}{2}$) cannot
exchange a pion ($I=1$).

Hypernuclear systems have been extensively studied experimentally for
a wide range of nuclei (from $^3_{\Lambda}$H to 
$^{208}_{~~\Lambda}{\mathrm Pb}$ \cite{hasegawa}) by 
employing hadronic processes such as stopped and low momentum kaon 
induced reactions, ${\mathrm A}(K,\pi) _{\Lambda} {\mathrm B}$, as well as
${\mathrm A}(\pi,K) _{\Lambda} {\mathrm B}$ reactions
(see Ref.~\cite{gibson} for a recent review of hypernuclear 
physics).  Nevertheless, since the different mechanisms are 
complementary, electromagnetic productions will, at some point, 
be required for a complete understanding of hypernuclear spectra. 

Several theoretical studies of hypernuclear  electromagnetic productions have
 been performed during the last few years  \cite{ben2,ben3}. 
The reactions $^{40}{\mathrm Ca}(\gamma,K^+)^{40}_{~\Lambda}{\mathrm K}$  and 
$^{208}{\mathrm Pb}(\gamma,K^+)^{208}_{~~\Lambda}{\mathrm Tl}$, for instance, have 
been calculated within the framework of a  {\it distorted wave impulse
approximation} (DWIA), where the interaction of 
the kaon with the final state has been included via a rather weak
optical potential derived from the elementary $KN$ amplitudes.
In contrast to the elementary processes, where both $S=J=0$ and 
$S=J=1$ transition terms contribute equally to the cross section,
the production from nuclei can eliminate $S=0$ or $S=1$ contributions
in certain transitions.  
The production of $\Sigma$ hypernuclei in reactions such as
$^{16}{\mathrm O}(\gamma, K^+)^{16}_{~\Sigma}{\mathrm N}$,
$^{40}{\mathrm Ca}(\gamma, K^+)^{40}_{~\Sigma}{\mathrm K}$, and
$^{208}{\mathrm Pb}(\gamma, K^+)^{208}_{~~\Sigma}{\mathrm Tl}$  \cite{ben3} has
also been calculated.

In this work we consider the reaction 
$^3{\mathrm He}(\gamma,K^+)^3_{\Lambda}{\mathrm H}$, i.e. 
the incoming real photon interacts with a nucleon (proton) in $^3$He 
creating a lambda which combines with the other two nucleons to form 
the bound hypertriton and a positively charged kaon which exits the nucleus. 
To our knowledge, no analysis has been made and no experimental data are 
available for this reaction. A recent calculation of Komarov
{\it et al.} \cite{komarov}, who studied the proton-nucleus collision
\begin{eqnarray}
  p + d \longrightarrow  K^+ +~^3_{\Lambda}{\mathrm H} 
\end{eqnarray}
estimated that at an incident proton energy $T_p$ = 1.13 - 3.0 GeV, 
the maximum differential cross section is well below 1 nb/sr, making 
experimental verification very difficult. It has been pointed out that 
this result is 50 times smaller than in the case of eta production 
through $p$-$d$ collisions. 

On the other hand, Tiator {\it et al.} \cite{tiator1} have estimated the 
differential cross section of eta photoproduction on $^3$He at 
$k=750$ MeV, close to threshold, to be around 
100 nb/sr at $\theta_{\mathrm c.m.} = 0^\circ$, with an expected decrease 
to 1 nb/sr at $\theta_{\mathrm c.m.} = 60^\circ$. Since the cross
section of elementary eta production is approximately 10 times larger 
than for the kaon, one would not expect a cross section 
larger than 10 nb/sr for kaon production on $^3$He. 

In this study we will evaluate the elementary operator for kaon
photoproduction between a realistic wave function of $^3$He, obtained as a 
solution of the Faddeev equations with the Reid soft core potential 
\cite{kim}, and the simple hypertriton wave function developed in Ref. 
\cite{congleton}. This simple hypertriton model used in our calculation 
\cite{congleton} has been adjusted to reproduce the experimental 
$\Lambda$-$d$ binding energy (0.13 $\pm$ 0.05 MeV) \cite{juric}, and it 
predicts the branching ratio \cite{keyes}
\begin{eqnarray}
  \label{branc}
  R &=& \frac{\Gamma (~^3_{\Lambda}{\mathrm H} \rightarrow \pi^- +~^3{\mathrm He}~)}{
              \Gamma (~^3_{\Lambda}{\mathrm H} \rightarrow \pi^- + {\mathrm all}~)}
    ~=~ 0.35 \pm 0.04 ~.
\end{eqnarray}
While we use this simple wave function for most calculations we
also perform comparisons with the correlated Faddeev wave function
of Ref.~\cite{miyagawa2} in order to assess the sensitivity of the cross
section predictions to different hypertriton descriptions.

In section \ref{section_wavefunction}, we briefly review the three-body wave 
functions used in our calculation along with some experimental facts on both 
$^3$He and the hypertriton. Section \ref{section_matrixelements} explains 
the matrix elements of the process. The present status of elementary models 
used in our calculation is briefly reviewed in section
\ref{section_elementary}. The results of our investigation are
presented and discussed in section \ref{section_results}. We
summarize our findings in section \ref{section_summary}.

\section{THE THREE-BODY WAVE FUNCTIONS}
\label{section_wavefunction}
Since both $^3$He and the hypertriton are three-body systems, we  
will describe the reaction using familiar three-body coordinates.
In the Jacobi representation, the three-body momentum coordinates for 
particles with momenta ${\vec k}_1$, ${\vec k}_2$, and ${\vec k}_3$, and 
masses $m_1$, $m_2$, and $m_3$, respectively, are given by 
\begin{eqnarray}
{\vec P} ~=~ {\vec k}_1 + {\vec k}_2 + {\vec k}_3 ~~,~~
{\vec p} ~=~ \frac{m_3 {\vec k}_2 - m_2 {\vec k}_3}{m_2 + m_3} ~~,~~
{\vec q} ~=~ \frac{(m_2 + m_3) {\vec k}_1 - m_1 ({\vec k}_2 + {\vec k}_3)}
{m_1 + m_2 + m_3} ~.
\label{jac1}
\end{eqnarray}
For the case of $^3$He, all constituents are assumed to have the same 
masses, and 
Eq.~(\ref{jac1}), in the center of momentum of the particles, reduces to 
\begin{eqnarray} 
{\vec P} ~=~ {\vec 0} ~~~,~~~
{\vec p} ~=~ {\textstyle \frac{1}{2}}\, ({\vec k}_2 - {\vec k}_3) ~~~,~~~
{\vec q} ~=~ {\vec k}_1 ~.
\label{jac2}
\end{eqnarray}
However, in the case of the hypertriton, the hyperon is clearly heavier than 
the proton or the neutron. Nevertheless, if we assume that the hyperon is  
particle 1, Eq.~(\ref{jac1}) may still be reduced to Eq.~(\ref{jac2}).

In Lovelace coordinates, the  expression corresponding to   
Eq.~(\ref{jac2}) is given by 
\begin{eqnarray}
  \label{lov1}
  {\vec P} ~=~ {\vec 0} ~~~,~~~
  {\vec p} ~=~ {\textstyle \frac{1}{2}}\, ({\vec k}_2 - {\vec k}_3) ~~~,~~~
  {\vec q} ~=~ -{\textstyle \frac{1}{2}}\sqrt{3}\, {\vec k}_1 ~.
\end{eqnarray}
Hence, the two coordinate systems differ only in the spectator coordinate 
by a factor of $-\frac{1}{2}\sqrt{3}$. Using the latter coordinate system
we will express the nuclear matrix element $T_{\mathrm fi}$ of the 
reaction in the lab frame as
\begin{eqnarray}
T_{\mathrm fi} &=& \langle~ ^3_{\Lambda}{\mathrm H}~|~ t^{\gamma p \rightarrow
K^+  \Lambda} ~|~ ^3{\mathrm He}~ \rangle ~, 
\label{tfi}
\end{eqnarray}
where the production operator, $t^{\gamma p \rightarrow K^+ \Lambda}$, 
is obtained from the elementary reaction. 

\subsection{The $^3$He Wave Function}

In our formalism, the three-body wave functions are expanded in orbital 
momentum, spin, and isospin of the pair (2,3) and the spectator (1) with 
the notation
\begin{eqnarray}
\Psi_{^3{\mathrm He}}({\vec p},{\vec q}\, ) &=& \sum_{\alpha} \phi_{\alpha}(p,q) 
~| (Ll){\cal L},(S{\textstyle \frac{1}{2}}){\cal S}, {\textstyle \frac{1}{2}} 
M \rangle ~| (T{\textstyle \frac{1}{2}}) {\textstyle \frac{1}{2}} M_t 
\rangle ~,
\label{hewf}
\end{eqnarray}
where $\phi_{\alpha}(p,q)$ stands for numerical solutions of 
Faddeev equations using the realistic nucleon-nucleon potential \cite{kim}. 

In Eq.~(\ref{hewf}) we have introduced $\alpha = \{ Ll{\cal L}S{\cal S}T \}$ 
to shorten the notation, where $L$, $S$, and $T$ are the total angular 
momentum, spin, and isospin of the pair (2,3), while for the spectator (1) the 
corresponding observables are labelled by $l$, $\frac{1}{2}$, 
and $\frac{1}{2}$, respectively. From now on we will use the Lovelace 
coordinates for the momenta of the pair and of the spectator. Their 
quantum numbers, along with the probabilities for the 11 partial waves, 
are listed in Table~\ref{phewf}. Clearly, most contributions will come 
from the first two partial waves (with a total probability of 88\%), 
which represent the $S$-waves with isospin 0 and 1, respectively.

\subsection{The Hypertriton}
The term ``hypertriton'' commonly refers to the bound state consisting 
of a proton, a neutron, and a lambda hyperon. Although a hypertriton 
consisting of a proton, a neutron, and a $\Sigma$ hyperon could exist, 
no experimental information is available at present \cite{sighyp}. 
The existing experimental information on the hypertriton is mostly 
from old bubble-chamber measurements \cite{oldi}. Table \ref{tb} compares
its properties with the triton and the deuteron.

Many models of the hypertriton have been developed using Faddeev 
equations \cite{miyagawa,miyagawa2,hypmod1}, the resonating group 
method \cite{hypmod2}, variational methods \cite{hypmod3}, and 
hyperspherical harmonics \cite{hypmod4}. We choose the simple model 
developed in Ref.~\cite{congleton}, which should be reliable enough
to obtain a first estimate for the photoproduction of the hypertriton.

In this model, the hypertriton is described by a deuteron and a lambda 
moving in an effective $\Lambda$-$d$ potential. The influence of
the lambda on the two nucleons is neglected, thus the nucleon part 
of the wave functions is exactly that of a free deuteron. We have also
neglected the $\Lambda N \rightarrow \Sigma N$ conversion, because the 
$\Sigma NN$ component in the hypertriton wave functions has been
calculated
to be very small. Using the phenomenological $YN$ potential developed in
Ref.~\cite{wycech}, the authors of Ref.~\cite{janus} found a probability
of only $0.36\%$ to have a $\Sigma NN$ component in the hypertriton.
This has been recently confirmed by the Bochum group. Using the Nijmegen 
$YN$ potential \cite{nijmegen1} and the Nijmegen93 $NN$ potential 
\cite{nijmegen2} they obtained a probability of $0.5\%$ \cite{miyagawa2}.
For other realistic $NN$ potentials the results are in the same range.

The effective $\Lambda$-$d$ potential is constructed as follows: First, 
a separable fit is performed to the $\Lambda N$ $S$-wave interaction 
given by the Nijmegen $YN$ soft-core potential \cite{nijmegen1}, which is 
then spin averaged over the $\Lambda N$ configurations found in the 
hypertriton. The $\Lambda N$ potential is summed over the 
two nucleons and averaged over the deuteron wave function. Finally, the 
resulting $\Lambda$-$d$ potential is fitted to a separable form, 
retaining only the $S$-wave part. The $\Lambda$-$d$ binding energy 
can then be reproduced by some fine tuning of the $\Lambda$-$d$ 
potential parameters.

With the notation of Eq.~(\ref{jac2}), the hypertriton wave function 
may be written as  
\begin{eqnarray}
\Psi_{^3_{\Lambda}{\mathrm H}}({\vec p},{\vec q}\, ) &=& 
\sum_{\alpha} \phi_{\alpha}(p,q)
~| (Ll){\cal L},(S{\textstyle \frac{1}{2}}){\cal S}, {\textstyle \frac{1}{2}} 
M \rangle  ~,
\label{hypwf}
\end{eqnarray}
where  $\phi_{\alpha}(p,q)$ is now simply given by the two 
separable wave functions of the deuteron and the lambda,
\begin{eqnarray}
\phi_{\alpha}(p,q) &=& \Psi^{(L)}_{d}(p)~\varphi_{\Lambda}(q) ~.
\end{eqnarray}
In Eq.~(\ref{hypwf}), we have dropped the isospin part of the 
wave function since the hypertriton has isospin 0. This argument
is based on the fact that only the $\Lambda np$ system appears 
to exist in nature, and that $\Lambda nn$ and $\Lambda pp$ bound 
systems have never been observed. Furthermore, it has been 
shown in Ref.~\cite{miyagawa2} that the states of the $\Lambda(\Sigma)NN$ 
system with quantum numbers $(T,J)$ different from $(0,\frac{1}{2})$ are 
not bound. Only the quantum numbers $\alpha = \{0001\frac{1}{2}0\}$ and 
$\{2021\frac{3}{2}0\}$ are non-zero in Eq.~(\ref{hypwf}). The probabilities 
for both partial waves are shown in the last column of Table~\ref{phewf}, 
where we have used the Paris potential for the deuteron part. It is clear 
that the two probabilities originate only from the deuteron, since the 
lambda part does not depend upon any of the quantum numbers.

The lambda part of the wave functions is found by solving the Schr\"odinger 
equation for a particle moving in the $\Lambda$-$d$ effective potential.
The solution is assumed to have the form  
\begin{eqnarray}
\label{lambdapart}
\varphi_{\Lambda}(q) &=& N(Q_{\Lambda})~ \frac{\exp [-(q/Q_{\Lambda})^2]}{
q^2 + \alpha^2} ~,
\end{eqnarray}
with $\alpha=(6.8 \pm 1.3) \times 10^{-2}~{\mathrm fm}^{-1}$, proportional 
to the square-root of the lambda binding energy, and the 
normalization factor 
\begin{eqnarray}
N(Q_{\Lambda}) &=& \left[~\frac{\pi}{4\alpha}~\left\{ {\mathrm cerfe}\left(
\frac{\sqrt{2} \alpha}{Q_{\Lambda}} \right)\left( 1+\frac{4\alpha^2}{
Q_{\Lambda}^2} \right) - \frac{2\alpha}{Q_{\Lambda}} \left(
\frac{2}{\pi} \right)^{\frac{1}{2}} \right\}~\right]^{-\frac{1}{2}} ~,
\end{eqnarray}
where
\begin{eqnarray}
  {\mathrm cerfe}(x) ~=~ {\mathrm exp}(x^2)\left[ 1- {\mathrm erf}(x)\right] ~~~,~~~
  {\mathrm erf}(x) ~=~ \frac{2}{\sqrt{\pi}}\int_0^x {\mathrm exp}(-t^2)~dt ~.
\end{eqnarray}
The author of Ref.~\cite{congleton} used the $\Lambda$-$d$ potential range 
$Q_{\Lambda} = 1.17~{\mathrm fm}^{-1}~(\sim \pm 10\%)$, leading 
to $[N(1.17)]^2 = 0.1039$. From Eq.~(\ref{lambdapart}) it is obvious that  
the lambda part of the hypertriton wave functions drops drastically 
as  function of the momentum $q$. It is also apparent that the most 
probable momenta of the lambda particle in the hypertriton are in 
the vicinity of 0.1 fm$^{-1}$. 

\section{THE MATRIX ELEMENTS}
\label{section_matrixelements}
\begin{figure}[htb]
\centerline{\psfig{figure=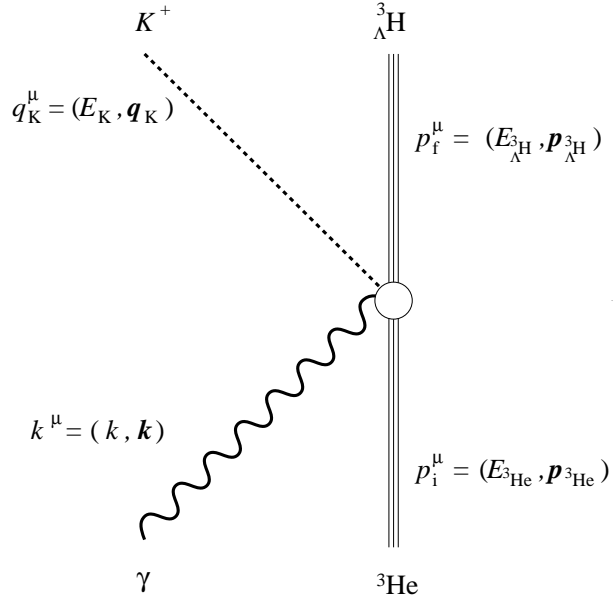,width=8cm}}
\caption{Kinematics for kaon photoproduction off $^3$He.}
\label{kin3he}
\end{figure}

\begin{figure}[tbp]
\centerline{\psfig{figure=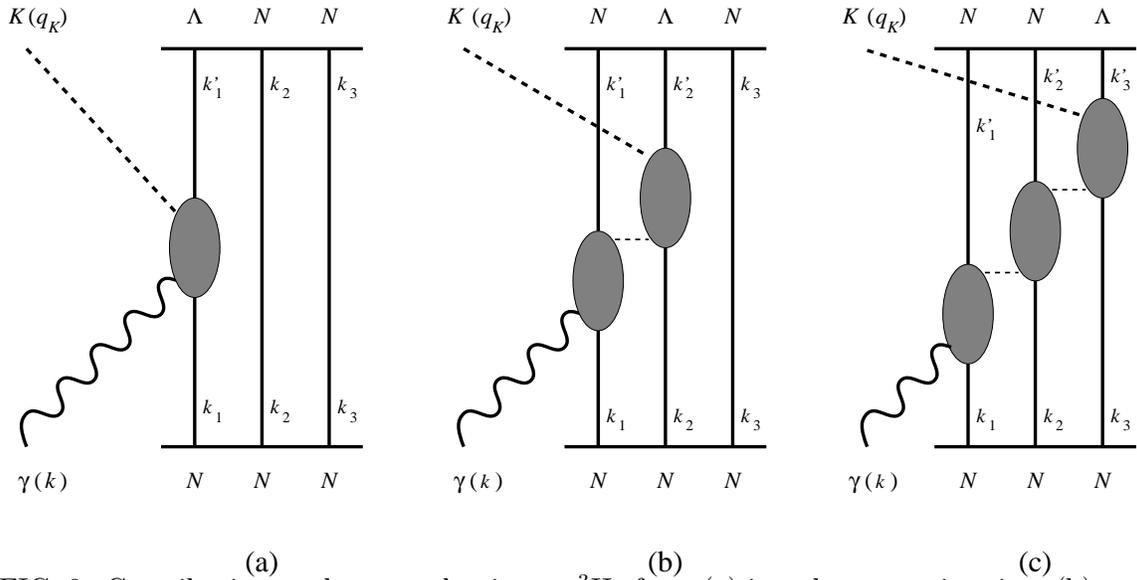,width=15cm}}
\caption{Contributions to kaon production on $^3$He from 
(a) impulse approximation, (b) and (c) back scattering terms
off two or three nucleons, respectively.}
\label{fpwia}
\end{figure}

Following the investigations of coherent pion photoproduction on $^3$He 
by Tiator {\it et al}.~\cite{tiator2,tiator2x,tiator3}, we calculate 
the reaction in momentum space. The Feynman diagram for photoproduction
of the hypertriton is depicted in Fig.~\ref{kin3he}, and the most 
important contributions to this process are shown in Fig.~\ref{fpwia}. 
For our present purpose, we will only consider the first diagram,  
corresponding to the impulse approximation, i.e.
the photon only interacts with one nucleon, while 
the other two nucleons of $^3$He act as spectators. We also neglect the
final state interaction (FSI) of the $K^+$ with the hypertriton. For
$^{12}{\mathrm C}(\gamma,K^+)^{12}_{~\Lambda}{\mathrm B}$, the $K^+$ FSI was
found to reduce the cross sections by 30\% \cite{ben3}, thus one would 
not expect FSI to affect our results by more than 5--10\%.

In the case of kaon photoproduction on the nucleus, the cross section
in the lab system can be written as 
\begin{eqnarray}
  \label{cssimple}
 \frac{d\sigma_{\mathrm T}}{d\Omega_K} &=& \frac{|{\vec q}_K^{\mathrm ~ c.m.}|}{|
{\vec k}^{\mathrm c.m.}|}~\frac{M_{\mathrm ^3He} 
E_{\mathrm ^3_{\Lambda}H}}{64\pi^2W^2}~\sum_{\epsilon}
\sum_{M,M'}~\left|T_{\mathrm fi}\right|^2 ~,
\end{eqnarray}
where the sums are over the photon polarization and over the initial
and final spin projections of the nucleus. 

The transition matrix elements can be expressed in 
terms of an integral over all internal momenta and states 
contributing to the process \cite{tiator2,tiator2x}, 
\begin{eqnarray}
T_{\mathrm fi} & = & \sqrt{3}
\int d^{3}{\vec p}~d^{3} {\vec q}
\Psi_{\mathrm ^{3}_{\Lambda}H}({\vec p}, {\vec q}\, ')~
t^{\gamma p \rightarrow K\Lambda}({\vec q}, {\vec Q})~ 
\Psi_{\mathrm ^{3}He}({\vec p}, {\vec q}\, ) \nonumber\\
&=& 2 \sqrt{6}~ 
\sum_{\scriptsize \left. \begin{array}{c} \alpha(Ll{\cal L}S{\cal S})\\
\alpha'(L'l'{\cal L}'S'{\cal S}')\\ n\lambda\Lambda m_{\Lambda} \end{array} 
\right.}
~\Biggl[~ i^{n}~\hat{n}\hat{\cal L}' \hat{\cal L}\hat{\cal S}' \hat{\cal S} 
\hat{\lambda}\hat{\Lambda}
~(-1)^{n+l'+l+L+S+{\cal L}+{\cal S}+M}~\delta_{LL'}~\delta_{SS'}~\delta_{T0}
\times \nonumber\\
&& \left( \begin{array}{rrr} \frac{1}{2} & \frac{1}{2} & \Lambda \\  
M' & -M & m_{\Lambda} \end{array} \right) 
 \left\{ \begin{array}{lll} {\cal L} & {\cal L}' & \lambda \\
l' & l & L \end{array} \right\}
\left\{ \begin{array}{lll}  {\cal S}' & {\cal S} & n \\  
\frac{1}{2} & \frac{1}{2} & S \end{array} \right\}
\left\{ \begin{array}{lll}  {\cal L} & {\cal S} & \frac{1}{2} \\  
{\cal L}' & {\cal S}' & \frac{1}{2}\\\lambda&n&\Lambda \end{array} \right\}
~I^{\alpha\alpha'}_{\lambda n \Lambda m_{\Lambda}}({\vec q},{\vec q}\, ') ~,
\label{tfi33}
\end{eqnarray}
with the four-dimensional integrals 
\begin{eqnarray}
I^{\alpha\alpha'}_{\lambda n \Lambda m_{\Lambda}}({\vec q},{\vec q}\, ') &=&
\int d^{3}{\vec q}~p^{2}dp~
\phi_{\alpha'}(p,q\, ')~ \phi_{\alpha}(p,q)~\times \nonumber\\
&& \Bigl[ \Bigl[{\bf Y}^{(l')}(\hat{{\vec q}\, '})\otimes
{\bf Y}^{(l)}(\hat{{\vec q}\, })\Bigr]^{(\lambda)} \otimes 
{\bf K}^{(n)}\Bigr]^{(\Lambda)}_{m_{\Lambda}}~
\end{eqnarray}
to be evaluated numerically. The factor of $\sqrt{3}$ on the right hand side
of Eq.~(\ref{tfi33}) comes from the antisymmetry of the initial state.
For the simple case of only $S$-state wave 
functions ($L'=l'={\cal L}'=0$, $S'=1$, ${\cal S}'=\frac{1}{2}$), 
Eq.~(\ref{tfi33}) reduces to 
\begin{eqnarray}
T_{\mathrm fi} & = &  \sqrt{\frac{6}{\pi}} 
~\sum_{\alpha {\alpha}'} ~\sum_{n \Lambda  m_{\Lambda}}
i^{n} \hat n \hat {{\cal L}} \hat {{\cal S}} \hat {{\cal S}'} \hat \Lambda 
(-)^{1 + n + {{\cal S}} + M} \delta_{SS'} \delta_{LL'} \delta_{T0} \times 
\nonumber\\ 
&& \left( \begin{array}{ccc} \frac{1}{2} & \frac{1}{2} & \Lambda\\
M' & -M & m_{\Lambda} \end{array} \right)
\left\{ \begin{array}{ccc} {{\cal S}} & {{\cal S}'} & n \\ \frac{1}{2} & 
\frac{1}{2} & 1 \end{array} \right\}
\left\{ \begin{array}{ccc} {{\cal L}} & {{\cal S}} & \frac{1}{2} \\ 
L & {{\cal S}'} & \frac{1}{2} \\
l & n & \Lambda \end{array} \right\} \times \nonumber\\
&& \int d^{3}{\vec q}~ p^{2}dp~ 
\varphi_{\Lambda}(q\, ')~ \Psi_{d}^{(L)}(p)~ \phi_{\alpha}(p,q)~ 
\left[ {\bf Y}^{(l)} (\hat{{\vec q}\, }) \otimes {\bf K}^{(n)} 
\right]^{(\Lambda )}_{m_{\Lambda}} ~, 
\label{tfi2}
\end{eqnarray}
In Eqs.~(\ref{tfi33}) and (\ref{tfi2}) we have used the 
Lovelace coordinate for the produced hyperon 
\begin{equation}
{\vec q}\, ' = {\vec q} - \frac{1}{\sqrt{3}} {\vec Q} ~,
\end{equation}
where ${\vec Q}$ is the momentum transfer to the nucleus, 
\begin{equation}
{\vec Q} = {\vec k} - {\vec q}_K ~.
\label{momtrans}
\end{equation}

Finally, the elementary production operator in Eq.~(\ref{tfi}), 
involving an invariant product between the photon polarization 
$\epsilon_{\mu}$ and the electromagnetic current $J_{\mu}$, 
has been decomposed into spin-independent and spin-dependent amplitudes
\begin{eqnarray}
t^{\gamma p \rightarrow K^+ \Lambda} &=& \epsilon_{\mu}~J^{\mu} \nonumber\\
&=& L + i {\vec{\sigma}} \cdot {\vec K} \nonumber\\
&=& \sum_{n=0,1} (-i)^{n~} \hat n ~\left[ 
{\mbox{\boldmath{$\sigma$}}}^{(n)} \otimes 
{\bf K}^{(n)} \right]^{(0)} ~,
\label{klndef}
\end{eqnarray}
with $\hat n = \sqrt{2n+1}$, ${\mbox{\boldmath{$\sigma$}}}^{(0)} = 1$, and 
${\bf K}^{(0)} = L$. The elementary production amplitudes $L$ and 
${\vec K}$ are calculated 
from the non-relativistic reduction of the elementary operator 
(see Appendix \ref{appnonrel}) and are given by 
\begin{eqnarray}
\label{lk1}
L & = & N~ \biggl\{ -{\cal{F}}_{14}~ {\vec p}_p
+ {\cal{F}}_{15}~ ({\vec q}_K - {\vec p}_p) \biggr\} \cdot ({\vec k} \times 
{\vec{\epsilon}}\, )\\
{\vec K} &=& -N~ \biggl[~ \biggl\{ {\cal{F}}_{1} + ({\cal{F}}_{14} - 
{\cal{F}}_{15})~ {\vec k}~ \cdot {\vec p}_p - {\cal{F}}_{15}~( |{\vec k}|^{2} 
- {\vec k} \cdot {\vec q}_K ) \biggr\}~{\vec{\epsilon}} \nonumber\\
&& +~ \biggl\{({\cal{F}}_{4} + {\cal{F}}_{5} + {\cal{F}}_{12} + 
{\cal{F}}_{13} - {\cal{F}}_{14} + {\cal{F}}_{15})~ {\vec p}_p \cdot 
{\vec{\epsilon}} - ({\cal{F}}_{5} + {\cal{F}}_{13} + {\cal{F}}_{15})~ 
{\vec q}_K \cdot {\vec{\epsilon}}~\biggr\}~ {\vec k} \nonumber\\
&& +~ \biggl\{ ({\cal{F}}_{8} + {\cal{F}}_{9} + {\cal{F}}_{12} + 
{\cal{F}}_{13})~ {\vec p}_p \cdot {\vec{\epsilon}} - 
( {\cal{F}}_{9} + {\cal{F}}_{13})~ {\vec q}_K 
\cdot {\vec{\epsilon}}~ \biggr\}~ {\vec p}_p \nonumber\\
&& +~ \biggl\{ - ({\cal{F}}_{12} + {\cal{F}}_{13})~ {\vec p}_p \cdot 
{\vec{\epsilon}} + {\cal{F}}_{13}~ {\vec q}_K \cdot 
{\vec{\epsilon}}~ \biggr\}~ {\vec q}_K ~ 
\biggr] ~,
\label{lk2}
\end{eqnarray}
where we have neglected small terms ${\cal F}_{16}$ - ${\cal F}_{20}$
in our non-relativistic approximation, and dropped all terms containing
$k^2$, ${\vec k} \cdot {\vec{\epsilon}}$, and 
$\epsilon_0$, since these terms will not contribute to photoproduction. 
It is easy to show that the omission of
${\cal F}_{16}$ - ${\cal F}_{20}$ will not destroy gauge invariance of 
the transition matrix. The analytical expressions of ${\cal F}_{1}$ - 
${\cal F}_{20}$ and $N$ are given in Appendix \ref{appnonrel}.

The tensor operators, $\left[ {\bf Y}^{(l)}(\hat{{\vec q}\, }) \otimes 
{\bf K}^{(n)} \right]^{(\Lambda )}_{m_{\Lambda}}$, which determine the 
specific nuclear transitions in the reaction, are given in Table~\ref{tensor}. 
In contrast to Ref.~\cite{tiator2}, the tensor operators in our case 
are simplified by the approximation that the hypertriton 
wave function only contains the partial wave with $l'=0$. However, for
future studies involving more advanced hypertriton wave functions 
\cite{miyagawa,miyagawa2}, the complete operator will be needed. For this
purpose, we have also derived the form of Eq.~(\ref{tfi2}) for the more
general case \cite{terry}.

Since both initial and final states of the nucleus are unpolarized, 
the sums over the spin projections can be performed by means of
\begin{eqnarray}
\sum_{M,M'}~\left|T_{\mathrm fi}\right|^2 &=& 
\sum_{\Lambda,m_{\Lambda}}~\left|T^{(\Lambda)}_{m_{\Lambda}}\right|^2 ~,
\end{eqnarray}
with 
\begin{eqnarray}
T^{(\Lambda)}_{m_{\Lambda}} &=& \sqrt{\frac{6}{\pi}}~
 \sum_{\alpha,{\alpha}',n}
\left[~ i^{n}~\hat{n}\hat{\cal L}\hat{S}' \hat{S}  
(-1)^{n+{\cal S}-\frac{1}{2}}
\left\{ \begin{array}{lll}  {\cal S}' & {\cal S} & n \\  
\frac{1}{2} & \frac{1}{2} & 1 \end{array} \right\}
\left\{ \begin{array}{lll}  {\cal L} & {\cal S} & \frac{1}{2} \\  
 L & {\cal S}' & \frac{1}{2}\\l&n&\Lambda \end{array} \right\}
 \delta_{LL'}\delta_{S1}\delta_{T0} ~ \times  \right. \nonumber\\
&& \int d^{3}{\vec q}~p^{2}dp~
\varphi_{\Lambda}(q\, ')~ \Psi_{d}^{(L)}(p)~\phi_{\alpha}(p,q)~
\Bigl[{\bf Y}^{(l)}(\hat{{\vec q}\, })\otimes
{\bf K}^{(n)}\Bigr]^{(\Lambda)}_{m_{\Lambda}} \Biggr] . ~
\label{tfi3}
\end{eqnarray}
Since the tensor ${\bf K}^{(n)}$ contains complicated functions of the 
integration variables ${\vec q}$ and $\hat{{\vec q}\, }=\Omega_q$, the 
integral 
in Eq.~(\ref{tfi3}) has to be performed numerically. It is appropriate 
to perform the overlap integration in $p$ first, because in the 
impulse approximation the tensor operator does not depend on
the relative pair momentum.

\section{Elementary Models}
\label{section_elementary}
Most current elementary models were developed to fit experimental 
data below 1.5 GeV. In recent analyses, only Refs.~\cite{bsaghai,williams} 
and the model of Ref.~\cite{terry} fit the photo- and electroproduction 
data up to 2.2 GeV. The recent analysis of
Ref.~\cite{bsaghai} gives a very comprehensive description of the 
elementary process. However, since this model incorporates
spin 5/2 resonances, the corresponding elementary operator is rather  
cumbersome for nuclear applications. Therefore, we will not include
this model in our calculations.
In Table \ref{newcc} we present the coupling constants 
for different models of the elementary reaction. We note that present
elementary models suffer from several fundamental uncertainties, such as 
the number of resonances to be included in view of the relatively high 
production threshold. For the sake of simplicity, current models
usually incorporate only few of them. Other complications arise from the 
extracted leading coupling constants, which are difficult to 
reconcile with the SU(3) predictions.

The elementary model developed in Ref.~\cite{terry} incorporates 
the intermediate $K^*$-exchange, the $N^*$ resonances $S_{11}$(1650) and 
$P_{11}$(1710) and, in addition, the $s$-channel $\Delta$ resonances
$S_{31}$(1900) and $P_{31}$(1910) for $K\Sigma$ photoproduction.
To achieve a reasonable $\chi^2$ for the experimental data in all six
isospin channels, Ref.~\cite{terry} introduced a hadronic form factor
of the form
\begin{eqnarray}
F_{\mathrm had}(\Lambda_{\mathrm c},t) & = & \frac{\Lambda_{\mathrm c}^2 - m_{K}^2}{ 
\Lambda_{\mathrm c}^2 - t} ~,
\end{eqnarray}
with $\Lambda_{\mathrm c}$ a cut-off parameter, which provides 
suppression at the higher energies and increases the leading coupling 
constants to values closer to the SU(3) prediction.

For the present purpose we will use the elementary models from 
Refs.~\cite{williams} and \cite{terry}, since we will investigate kaon
photoproduction on $^3$He with simple elementary operators giving a  
reasonable description at relatively high energies.

\section{Results and Discussion}
\label{section_results}
\begin{figure}[tbp]
\centerline{\psfig{figure=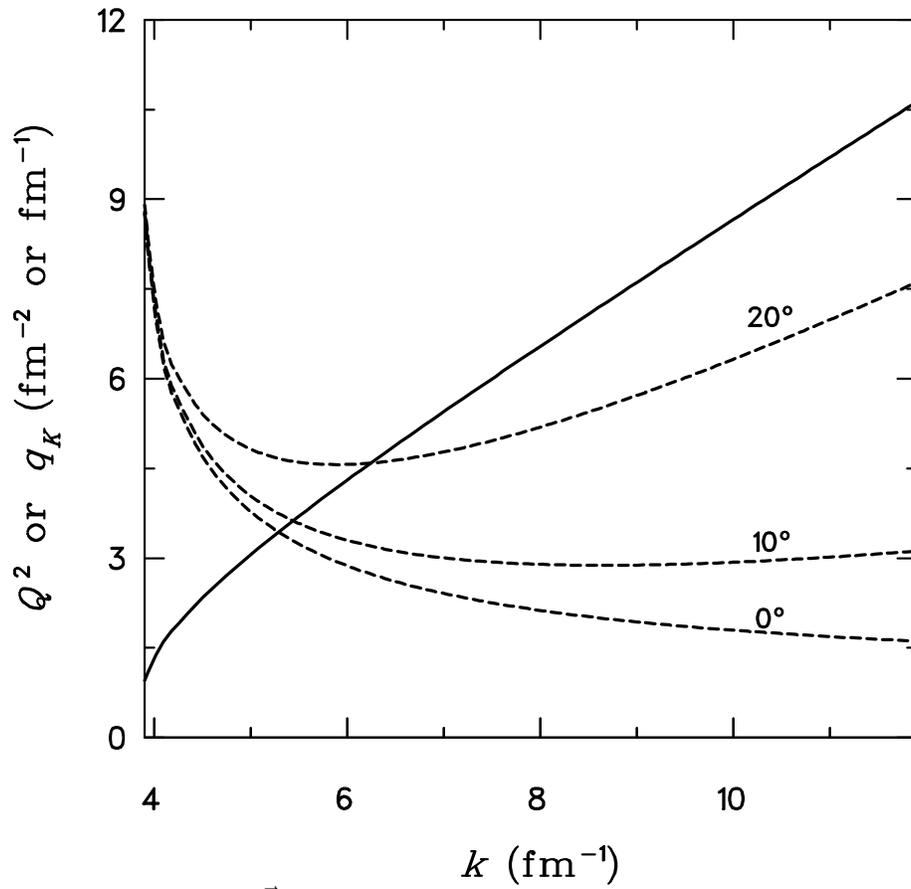,width=12cm}}
\caption{Momentum transfer ${\vec Q}^{\, 2}$ (dashed curves) and kaon 
         momentum $q_K=|{\vec q}_K|$ 
         (solid curves) as a function of the photon laboratory energy
         for 3 different kaon angles.}
\label{q2vsk}
\end{figure}

Both kaon photo- and electroproduction off $^3$He can be analyzed using
the formalism introduced in the preceding sections. However, as a first
step, we will concentrate on photoproduction, since this process 
is simpler than the virtual case. We first search for kinematical 
situations where the cross section will be maximum by 
inspecting the elementary process. Since the cross section tends to 
increase with the excitation energy, we decided to investigate 
the observables at energies $k=1.4-2.2$ GeV, where we expect 
the reaction rates to be reasonably high. It is also well known that the 
maximum cross section can be achieved at minimum momentum transfer, 
i.e. at forward angles. However, even in this region the corresponding 
momentum transfers are already large, i.e. $Q\simeq 1.29 - 1.54$ fm$^{-1}$.
Since the momentum transfer increases rapidly with the scattering 
angle (see Fig.~\ref{q2vsk}), the nuclear form factor will strongly 
suppress the cross sections at larger angles.

The isospin formalism has to assure that  $K^+\Lambda$ production 
occurs only on protons in $^3$He. Indeed 
the matrix element contains a delta function $\delta_{T0}$ 
[see Eq.~(\ref{tfi3})] which excludes the
contributions coming from the proton-proton pair in $^3$He, i.e. the
production on the neutron. In the S-wave approximation, where both
$T=0$ and $T=1$ partial waves exist in $^3$He, i.e.
\begin{eqnarray}
  \label{swapprox}
  \Psi_{^3{\mathrm He}}({\vec p},{\vec q}\, ) &=& \frac{1}{\sqrt{2}}~ \left[
  ~\phi_1(p,q)~|T=0,S=1\rangle +\phi_2(p,q)~|T=1,S=0\rangle~ \right] ~,
\end{eqnarray}
but only the partial wave with $T=0$ exists in the hypertriton, the delta
function yields a reduction in the cross section by a factor of two,
if we assume that both $\phi_1(p,q)$ and $\phi_2(p,q)$ are normalized
to 1. In  realistic wave functions, however, it is the sum of all partial 
probabilities that is normalized to 1 (see Table \ref{phewf}).

As a check of our calculations and computer codes, we compare the full 
result with two simple approximations. First, we reduce the cross
section by allowing only $S$--waves to contribute to the amplitudes
in Eq.~(\ref{tfi3}). This approximation should be reasonable because, as
shown in 
Table~\ref{phewf}, contributions from other partial waves are small.
In this approximation, Eq.~(\ref{cssimple}) reduces to
\begin{eqnarray}
  \label{cssimple2}
 \frac{d\sigma_{\mathrm T}}{d\Omega_K} &=& 
\frac{|{\vec q}_K^{\mathrm ~ c.m.}|_{\mathrm ^3He}
}{|{\vec k}^{\mathrm c.m.}|_{\mathrm ^3He}}~\frac{M_{\mathrm ^3He} 
E_{\mathrm ^3_{\Lambda}H}}{32\pi^2W_{\mathrm ^3He}^2}~\sum_{\epsilon}~\left( 
{\textstyle \frac{3}{2}}~|{\tilde L}|^2 
+ {\textstyle \frac{1}{6}}~|{\tilde{{\vec K}\, }}|^2~\right) ~,
\end{eqnarray}
where
\begin{eqnarray}
  \label{ltilde}
  {\tilde L} &=& \frac{1}{4\pi} 
\int d^{3}{\vec q}~p^{2}dp~
\varphi_{\Lambda}(q\, ')~\Psi_{d}^{(0)}(p)~\phi_{1}(p,q)~L({\vec q},
{\vec q}\, ') ~, ~~~~
\end{eqnarray}
and
\begin{eqnarray}
  \label{ktilde}
  {\tilde{{\vec K}\, }} &=& \frac{1}{4\pi} 
\int d^{3}{\vec q}~p^{2}dp~
\varphi_{\Lambda}(q\, ')~\Psi_{d}^{(0)}(p)~\phi_{1}(p,q)~{\vec K}({\vec q},
{\vec q}\, ') ~. ~~~~
\end{eqnarray}
Note that in the integrals above we have already excluded 
the contribution from the S-wave with $T=1$ ($\phi_{2}$) and assumed that
$\phi_1(p,q)$ is normalized to unity.

Apart from the factors of $\frac{3}{2}$ and $\frac{1}{6}$ in front of the 
amplitudes $|{\tilde{{\vec L}\, }}|^2$ and $|{\tilde{{\vec K}\, }}|^2$, 
Eq.~(\ref{cssimple2}) is similar to the cross section for elementary
photoproduction. We recall that in this case the cross section is given by
\begin{eqnarray}
  \label{csnucleon}
 \left( \frac{d\sigma_{\mathrm T}}{d\Omega_K} \right)_{\mathrm proton}
&=& ~\frac{|{\vec q}_K^{\mathrm ~ c.m.}|_p}{|{\vec k}^{\mathrm c.m.}|_p}~
\frac{m_p E_{\Lambda}}{32\pi^2W_p^2}~\sum_{\epsilon}
~\left( ~|L|^2 + |{\vec K}|^2~\right) ~.
\end{eqnarray}
Note that in Eqs.~(\ref{cssimple2}) and (\ref{csnucleon}) extra subscripts
have been added in order to distinguish between the kinematic variables
for the proton and for $^3$He.

At $k=1.8$ GeV, we found\footnote{This situation is different in pion 
photoproduction, where the $L$ and ${\vec K}$ amplitudes are  comparable.}
 that $|L|^2 \ll |{\vec K}|^2$ and 
$|{\tilde L}|^2 \ll |{\tilde{{\vec K}\, }}|^2$. Therefore, to a good 
approximation, the ratio of the cross section for $^3$He to the 
elementary cross section is given by
\begin{eqnarray}
  \label{ratio}
  \frac{d\sigma_{\mathrm T} ({\mathrm ^3He})}{d\sigma_{\mathrm T} (p)} &~\approx~& 
 \frac{|{\vec q}_K^{\mathrm ~ c.m.}|_{\mathrm ^3He}}{|{\vec k}^{\mathrm c.m.}|_{\mathrm ^3He}}~
 \frac{|{\vec k}^{\mathrm c.m.}|_p}{|{\vec q}_K^{\mathrm ~ c.m.}|_p}~
 \frac{M_{\mathrm ^3He} E_{\mathrm ^3_{\Lambda}H} W_p^2}{m_p E_{\Lambda}
  W_{\mathrm ^3He}^2} ~\frac{1}{6}~
  \frac{\displaystyle \sum_\epsilon |{\tilde {{\vec K}\, }}|^2}{\displaystyle 
 \sum_\epsilon |{\vec K}|^2} \nonumber\\
 &~\approx~& 1.8 \times \frac{1}{6} \times 19.6 \times 10^{-3} \nonumber\\
 &~\approx~& 5.9 \times 10^{-3} ~,
\end{eqnarray}
where we have used the realistic $^3$He wave function along with the simple 
model of the hypertriton in Eq.~(\ref{ktilde}).

At this energy, the elementary reaction model of Ref.~\cite{williams} yields 
a maximum cross section of about 500 nb/sr. As a consequence we 
can expect a cross section of about 3 nb/sr for photoproduction
at $k=1.8$ GeV.

As a second approximation, we consider the struck nucleon inside $^3$He  
as having a fixed momentum \cite{kamalov,sabit}. Therefore, the 
${\vec K}$ amplitude in Eq.~(\ref{ktilde}) can be factored out of the integral 
\begin{eqnarray}
  {\tilde {{\vec K}\, }} &=& {\vec K}(Q) ~ F(Q) ~,
\end{eqnarray}
and the cross
section off $^3$He may be written as 
\begin{eqnarray}
  \label{msabit}
 \frac{d\sigma_{\mathrm T}}{d\Omega_K} &=&{\textstyle \frac{1}{6}}~ W_A^2~|F(Q)|^2 
~\left( \frac{d\sigma_{\mathrm T}}{d\Omega_K}\right)_{\mathrm proton} ~,
\end{eqnarray}
where ${\vec K}$ now only depends on the momentum transfer and 
 the nuclear form factor\footnote{Note that we use the Jacobi coordinate
system for convenience.}
\begin{eqnarray}
  \label{efq1}
F(Q) &=& \int~d^3{\vec q}~d^3{\vec p}~~
\Psi_{^3_{\Lambda}{\mathrm \!H}}({\vec p},{\vec q}+{\textstyle \frac{2}{3}}
{\vec Q})~~ \Psi_{^3{\mathrm He}}({\vec p},{\vec q}\, ) \nonumber\\
     &=& \frac{1}{4\pi} \int d^{3}{\vec q}~p^{2}dp~ 
         \varphi_{\Lambda}(q\, ')~\Psi_{d}^{(0)}(p)~\phi_{1}(p,q) \nonumber\\
     &=& 0.69~{\mathrm exp}\left(-{\textstyle \frac{4}{9}} b_{\Lambda}^2 Q^2\right)
 \times \nonumber\\ 
  &&  \int q^2dq~d\cos\theta ~ \frac{{\mathrm exp}\left\{ -\left[ (b_{\Lambda}^2+
\frac{3}{4}b^2)q^2+\frac{4}{3}b_{\Lambda}^2qQ\cos\theta \right] 
\right\} }{q^2+\frac{4}{9}Q^2+\frac{4}{3}qQ\cos\theta+\alpha^2} ~,
\end{eqnarray}
where $Q=|{\vec Q}|$, $q=|{\vec q}\, |$, and 
$q'=|{\vec q}+\frac{2}{3}{\vec Q}|$. The kinematical factor in 
Eq.~(\ref{msabit}) is given by Eq.~(\ref{ratio}), i.e.
\begin{eqnarray}
  \label{kinfac}
  W_A &=& \sqrt{ \frac{|{\vec q}_K^{\mathrm ~ c.m.}
 |_{\mathrm ^3He}}{|{\vec k}^{\mathrm c.m.}|_{\mathrm ^3He}}~
 \frac{|{\vec k}^{\mathrm c.m.}|_p}{|{\vec q}_K^{\mathrm ~ c.m.}|_p}~
 \frac{M_{\mathrm ^3He} E_{\mathrm ^3_{\Lambda}H} W_p^2}{m_p E_{\Lambda}
  W_{\mathrm ^3He}^2} } ~
\end{eqnarray}

To obtain the last part of Eq.~(\ref{efq1}), we have parametrized the  $^3$He 
and deuteron wave functions by Gaussians, 
\begin{eqnarray}
  \label{parametrize1}
   \phi_{1}(p,q) &=& \left( \frac{48\sqrt{3}~b^6}{\pi}
\right)^{1/2}~ {\mathrm exp}\left[ -b^2\left( p^2+{\textstyle \frac{3}{4}}
q^2\right)\right] ~,
\end{eqnarray}
and
\begin{eqnarray}
  \label{parametrize2}
  \Psi_d^{(l)}(p) &=& \left( 8\sqrt{\frac{2}{\pi}}~b_d^3 \right)^{1/2}~ 
{\mathrm exp}\left[ -b_d^2~ p^2 \right] ~,
\end{eqnarray}
with $b=1.65$ fm, $b_d=1.58$ fm, and Eq.~(\ref{lambdapart})
for the lambda part of the hypertriton wave function.

The factor of $\frac{1}{6}$, appearing in Eq.~(\ref{cssimple2}), is the 
result of a specific nuclear transition in the process (recall that only 
the state with $L=l={\cal L}=T=0$, $S=1$, and ${\cal S}=\frac{1}{2}$ 
contributes) and the normalization of nuclear wave functions. 
Along with the fact that $|L|^2\ll |{\vec K}|^2$ in 
elementary kaon production, it leads to a large reduction of the  
cross section. 
We note that if the hypertriton would have an excited state with $J=3/2$,
this state would be preferentially formed by a ratio of 8:1 with respect
to the $J=1/2$ ground state. However, no excited state of the hypertriton is
known, the $J=3/2$ state is therefore unbound and lies in the $K \Lambda$
quasifree production continuum.

\begin{figure}[tbp]
\centerline{\psfig{figure=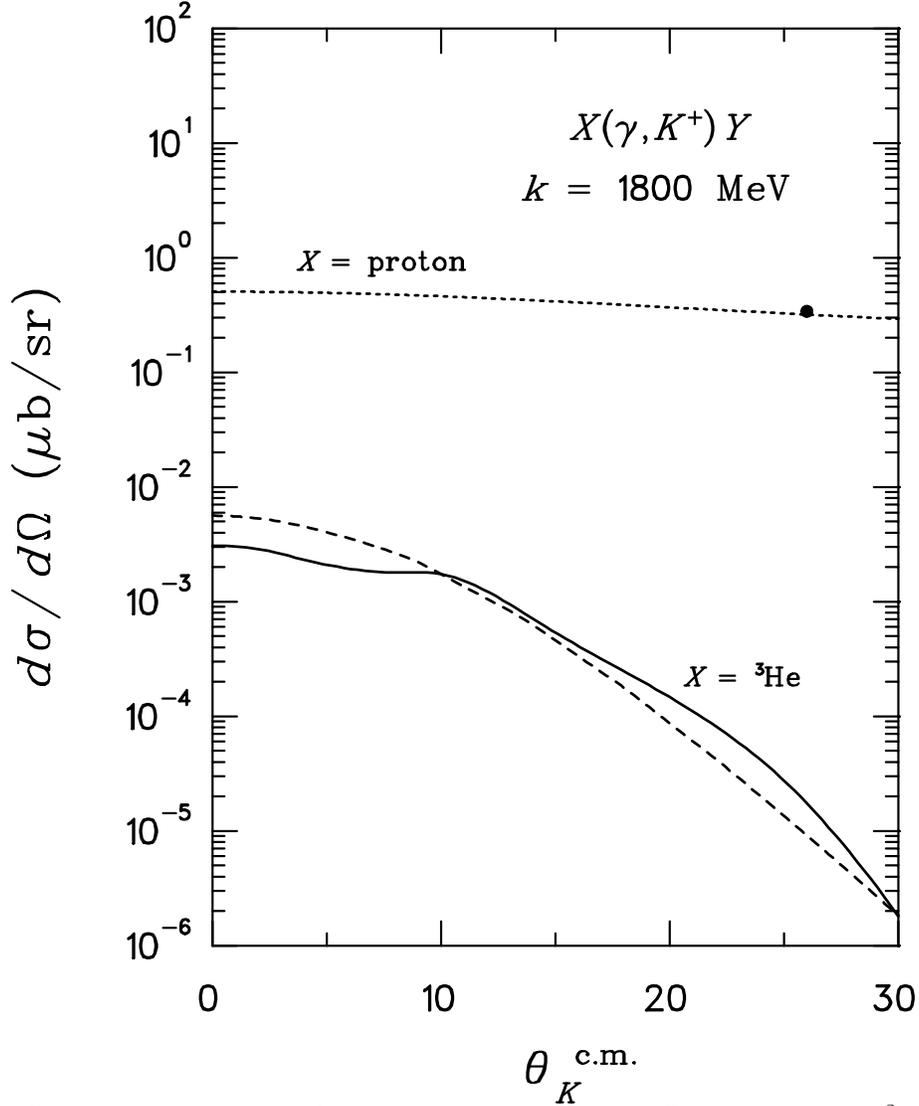,width=12cm}}
\caption{Differential cross section for kaon photoproduction off the proton
 and $^3$He as function of kaon angle. The elementary reaction (dotted 
 line) is taken from  Ref.~\protect\cite{williams} and the corresponding 
 experimental datum is from Ref.~\protect\cite{fe}. The dashed line shows the
 approximation for production off $^3$He calculated from Eq.~(\ref{msabit}), 
 the solid line represents the exact calculation using $S$-waves.}
\label{ffq2}
\end{figure}

Using Eq.~(\ref{efq1}) it can be shown that the nuclear form factor 
reduces the reaction cross section of Eq.~(\ref{msabit}) by more than 
a factor of 25. The result is displayed in Fig.~\ref{ffq2}. 
The nuclear cross section at forward angles is smaller 
than that of elementary kaon production by two orders of magnitude.
As $\theta_K^{\mathrm c.m.}$ increases, the cross section drops quickly, 
since the nuclear momentum transfer increases as function of 
$\theta_K^{\mathrm c.m.}$ (see Fig.~\ref{q2vsk}). 

Figure \ref{ffq2} also shows the significant difference between the cross
sections calculated with the approximation of Eq.~(\ref{msabit}) and 
the full result obtained from Eq.~(\ref{cssimple}). This discrepancy is  
due to the ``factorization'' approximation, since in the full 
calculation both spin-independent and spin-dependent amplitudes 
are integrated over the internal momentum and weighted by the two wave 
functions. Furthermore, in
Eq.~(\ref{msabit}) we use simple parametrizations for both $^3$He and deuteron
wave functions [Eqs.~(\ref{parametrize1}) and (\ref{parametrize2})].

The cross section 
for kaon photoproduction is in fact very small, of the order of several 
nanobarns at most, and even smaller for larger kaon angles. This is in 
contrast 
to other hypernuclear reactions, e.g. in the case of $^{16}_{~\Lambda}$N 
and $^{40}_{~\Lambda}$K  production, where cross sections of the order 
of several hundreds nanobarns have been predicted \cite{ben3}. The 
underlying reason is the lack of high momentum components in the
$^3_\Lambda{\mathrm H}$ wave function. Since the momentum transfers are 
high, the lambda momentum is high as well, which inhibits hypernuclear 
formation. Nevertheless, the electromagnetic production of the hypertriton 
has to be compared to the production with strong probes, e.g.
\begin{eqnarray}
  \label{pd-proc}
  p ~+~ d ~\longrightarrow~ K^+ ~+~ ^3_{\Lambda}{\mathrm H} ~.
\end{eqnarray}
As stated before, Komarov {\it et al.} \cite{komarov} have predicted 
cross sections smaller than 1 nb/sr for the same hypertriton wave 
function \cite{congleton} as in our work. Their calculation predicts a 
cross section with a maximum at an incident proton kinetic energy 
of 1.35 GeV and an emission angle $\theta_K^{\mathrm c.m.}=180^\circ$.

A sufficient number of integration points is found to be essential 
for the stability of our results. In contrast to pion photoproduction, 
where both initial and final states  have the same wave function, 
the hypertriton  wave function in momentum space drops faster than 
in the case of $^3$He one. Former studies of pion 
photoproduction off $^3$He \cite{tiator2} used a four-dimensional 
integration with $5\times 5$ grid points for the angular integration.
Such an integration was found to be insufficient for our purpose.  
As shown in Fig.~\ref{q2vsk}, the momentum transfer at the energy of 
interest and large kaon angles, increases quickly as a function 
of excitation energy, thus strongly suppressing the cross section 
at the corresponding angle. As a consequence, a relatively small grid
size is required to obtain accurate results.
To investigate the sensitivity of the integration to the grid number ($n$),
we carried out the calculation of the angular integration as a function
of $n$ up to $n=50$. It is found that the integrations with $n=5$ and 
$n=10$ yield very different cross sections 
with a discrepancy by more than 100\% at the forward angles, and start to 
fluctuate as the angle increases. Only at $n\geq 20$ the 
integration begins to become stable. Therefore, we have performed the 
calculations with $20\times 20$ angular grid points. 
For the integrations over the momenta $p$ and $q$, we follow the 
work of Tiator {\it et al.} \cite{tiator2,tiator2x,tiator3}, i.e. using 
$n_p ({\mathrm max}) = 14$ and $n_q({\mathrm max}) = 24$. Since the result 
using $n_q=20$ does not significantly differ from that one with $n_q=24$,
we have eventually carried out an integration over 
$14\times 20 \times 20 \times 20$ grid points. 

\begin{figure}[tbp]
\centerline{\psfig{figure=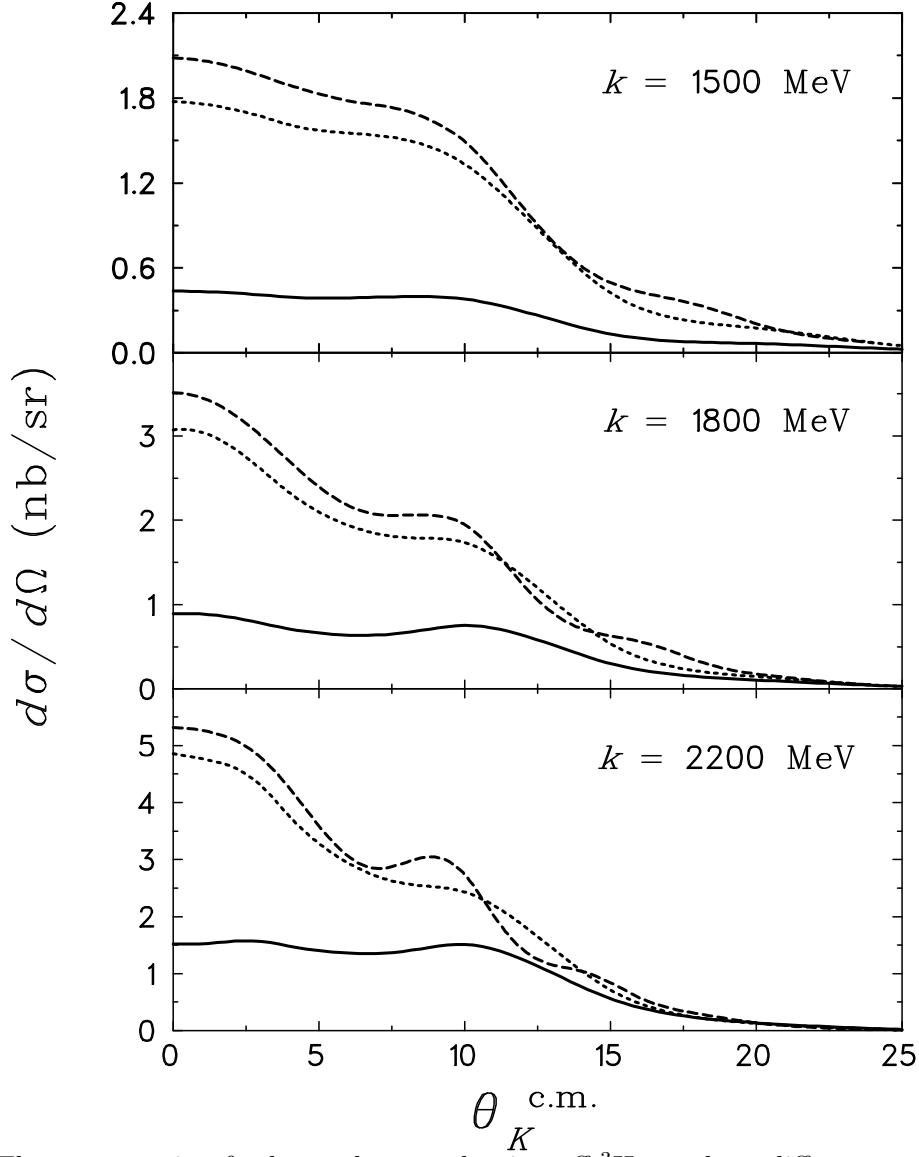,width=12cm}}
\caption{The cross section for kaon photoproduction off $^3$He at three
 different excitation energies. The dotted curves are obtained from the 
 the calculation with $S$--waves only and the simple hypertriton wave
function, the dashed curves are obtained with $S$--waves only
and the correlated Faddeev wave function of Ref.~\protect\cite{miyagawa2},
while the solid curves show the result after using all of the partial 
waves and the simple hypertriton wave function.}
\label{fvarwave}
\end{figure}

A surprising result is shown in Fig.~\ref{fvarwave}. In contrast to our 
previous conjecture that the contribution should mostly come from
$S$--waves (as in the case of pion photoproduction \cite{tiator2}), 
the higher partial waves further reduce the cross section by a factor 
of more than three. The reason can be traced back to Table~\ref{phewf}.
The three Kronecker delta functions in Eq.~(\ref{tfi3}) yield selection
rules which allow a transition from an initial state with $\alpha =
1$ or 8 to the final state with $\alpha ' =1$, and from the states with 
$\alpha = 3$ or 7 to the state with $\alpha '=7$ only. The transitions 
from $\alpha=7$ to $\alpha ' =7$ as well as from $\alpha=3$ to 
$\alpha ' =7$ are negligibly small. However, the transition 
from $\alpha=8$ to $\alpha ' = 1$ may not be neglected, since $\alpha '=1$ 
(with the probability of about 94\%) is the most likely state in the 
hypertriton. In the case of pion production this transition is negligible
mainly because the $S$-waves with $\alpha=1$ and 2 (with probabilities 
of 44.3\% and 43.7\%, respectively) dominate all transitions.
We also note that the angular momentum part of the tensor amplitude 
in Eq.~(\ref{tfi3}) gives a considerable contribution for both leading 
transitions ($\alpha =1,8$ to $\alpha '=1$). Hence, in 
the following calculations we always include the complete set of partial 
waves ($\alpha=1,3,7,8$ and $\alpha '=1,7$). In comparison, the 
higher partial waves in pion photo- and electroproduction 
decrease the cross section by at most 15\% and 20\%, respectively.

Since the $(\gamma,K)$ process is a high momentum transfer process and the 
simple analytical hypertriton wave function used until now contains no 
short-range correlations we also show in Fig.~\ref{fvarwave} a comparison
with the correlated three-body wave function of Ref.~\cite{miyagawa2}
that includes proper short-range behavior. While the cross section
obtained with the Faddeev wave function shows more structures the
differences are only of order 10-20$\%$. The absence of short-range
correlations in the simple hypertriton model does not become obvious
until momentum transfers outside the range considered here. We therefore
continue using the simple hypertriton wave function for the following 
calculations as well.

The small size of the cross section obtained here raises the question
of the possible significance of two-step processes, such as
$\gamma + p \rightarrow p + \pi^0 \rightarrow K^+ \Lambda$.
Two-step processes were studied in Ref.~\cite{kamalov95} for pion
photoproduction on $^3$He and found to be significant only at
much larger $Q^2$ compared to this study. Ref.\cite{fix97} also 
included  these processes in $\eta$ photoproduction on the deuteron
and found only small effects.  However, a future investigation would
have to study this question in more detail for kaon photoproduction,
including these effects here would go beyond the realm of this work.

\begin{figure}
\centerline{\psfig{figure=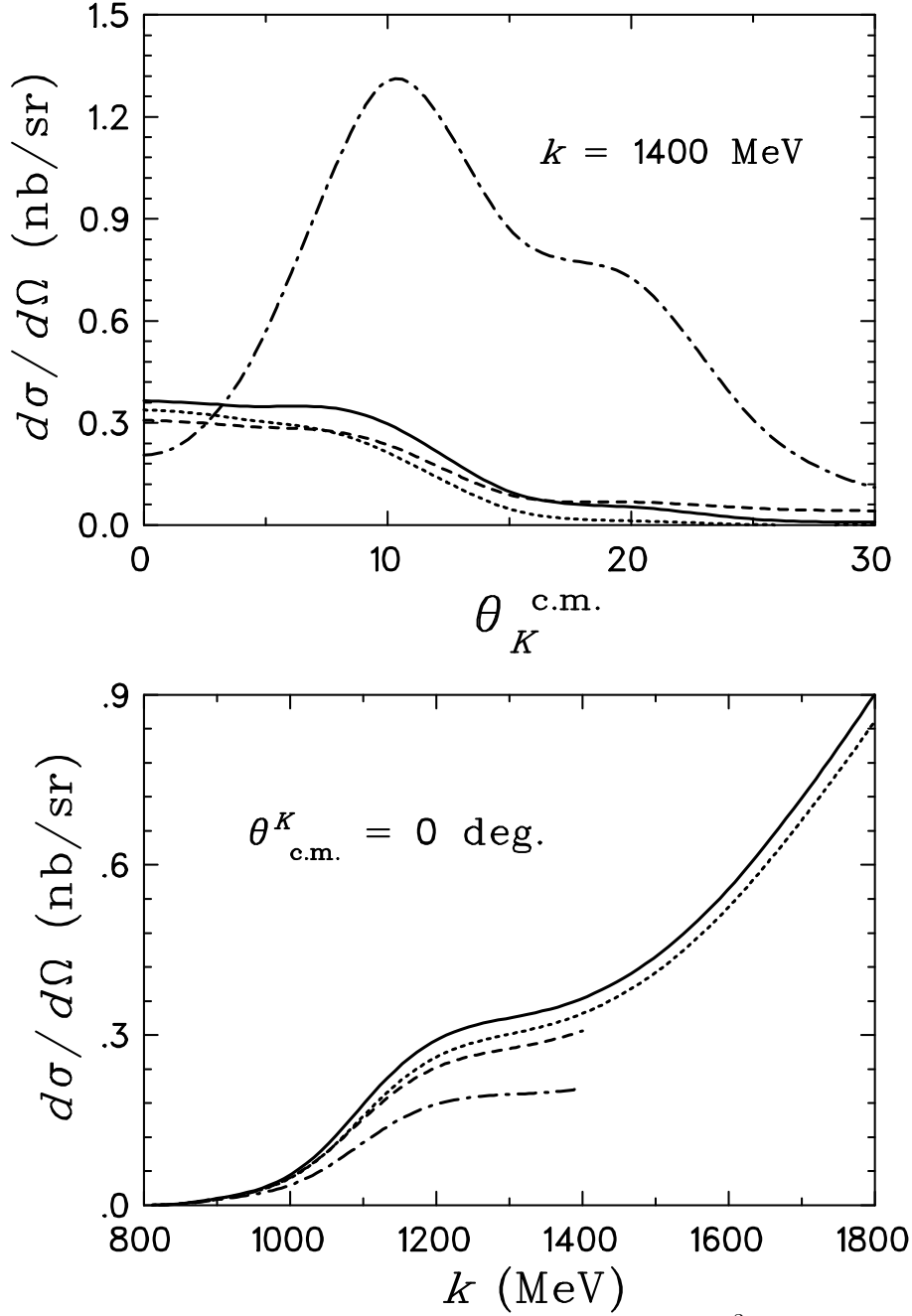,width=12cm}}
\caption{Differential cross sections for kaon photoproduction off $^3$He
predicted by different elementary models. The dash-dotted and the dashed 
curves are the predictions of the elementary models of 
Refs.~\protect\cite{adel2} and \protect\cite{adel1} (set I and set II 
of Table \ref{newcc}, respectively), the dotted curve is obtained with the 
coupling constants of set IV \protect\cite{terry}, and the solid curve is 
the result for using the model of Ref.~\protect\cite{williams} (set III). 
The first two models (dashed and dash-dotted curves) fit the kaon data only 
up to 1.4 GeV. All curves are calculated by using all partial waves.}
\label{fvarmod}
\end{figure}

In Fig.~\ref{fvarmod}, we compare the cross sections predicted by 
different elementary models. Except for the model of Ref.~\cite{adel2},
all models produce similar cross sections at $k \leq 1.4$ GeV.
The different feature predicted by the model of Ref.~\cite{adel2} can 
be understood from the fact that this model overestimates  
the experimental data at $k\geq 1.3$ GeV and $0^\circ < \theta_K^{\mathrm c.m.} 
< 30^\circ$ by about 40\%. The elementary model developed 
in Ref.~\cite{terry} and that of Ref.~\cite{williams} 
are preferred, since both explain the elementary 
photoproduction data up to 2.2 GeV, where reasonable cross sections off 
$^3$He might be expected. However, for the sake of simplicity, we will
use the model of Ref.~\cite{williams} in the subsequent calculations.

We have investigated the contribution of non-localities generated
by Fermi motion in the initial and final nuclei. As in former studies 
\cite{ben3,tiator2}, an exact treatment of Fermi motion is included 
in the integrations over the  wave functions in Eq.~(\ref{tfi3}), while 
a local approximation can be carried out by freezing the operator 
at an average nucleon momentum 
\begin{eqnarray}
\label{local}
\langle {\vec k}_1 \rangle &=& -\kappa \frac{A-1}{2A}{\vec Q} ~, 
\end{eqnarray} 
where in this case, $A=3$. For $\kappa =0$, Eq.~(\ref{local}) 
corresponds to the ``frozen nucleon'' approximation, whereas  
$\kappa =1$  yields the average momentum approximation. 
The latter case has been shown to yield satisfactory results 
for pion photoproduction in the $s$- and $p$-shells \cite{tiator3}. 
Furthermore, as shown in Refs.~\cite{tiator2,tia_th} in the case of 
pion photoproduction, Fermi motion can be approximated by choosing 
$\kappa=1$. This approximation can reproduce the exact cross section 
to within an accuracy of 7\% \cite{macdonald}.

\begin{figure}[tbp]
\centerline{\psfig{figure=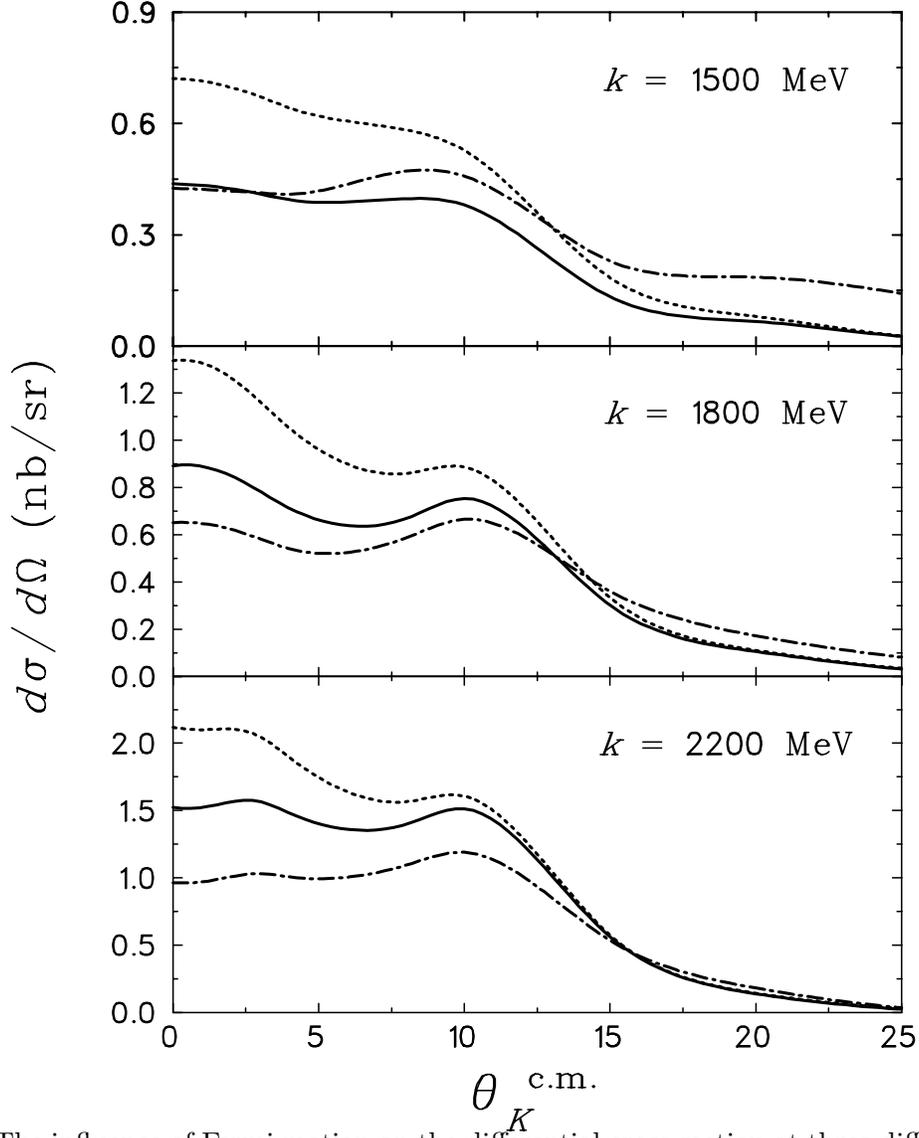,width=12cm}}
\caption{The influence of Fermi motion on the differential cross section
 at three different photon energies. The dash-dotted curve is the 
 ``frozen nucleon'' approximation ($\langle {\vec k}_1 \rangle = 0$),
 the dotted curve is obtained with an average momentum of 
 $\langle {\vec k}_1 \rangle = -\frac{1}{3}{\vec Q}$, while the solid
 curve shows the exact result.}
\label{fvarfm}
\end{figure}

Figure~\ref{fvarfm} compares the cross sections calculated in the 
two approximations with the exact calculation. A systematic discrepancy 
between the calculation with Fermi motion and the one with the average 
momentum approximation appears at all energies. Unlike in pion 
photoproduction, the average momentum approximation cannot simulate 
Fermi motion in kaon photoproduction, and the 
discrepancies between the different methods, especially near forward 
angles, are too significant to be neglected. Based on this result,  
all further calculations are performed considering Fermi motion exactly.

\begin{figure}[t]
\centerline{\psfig{figure=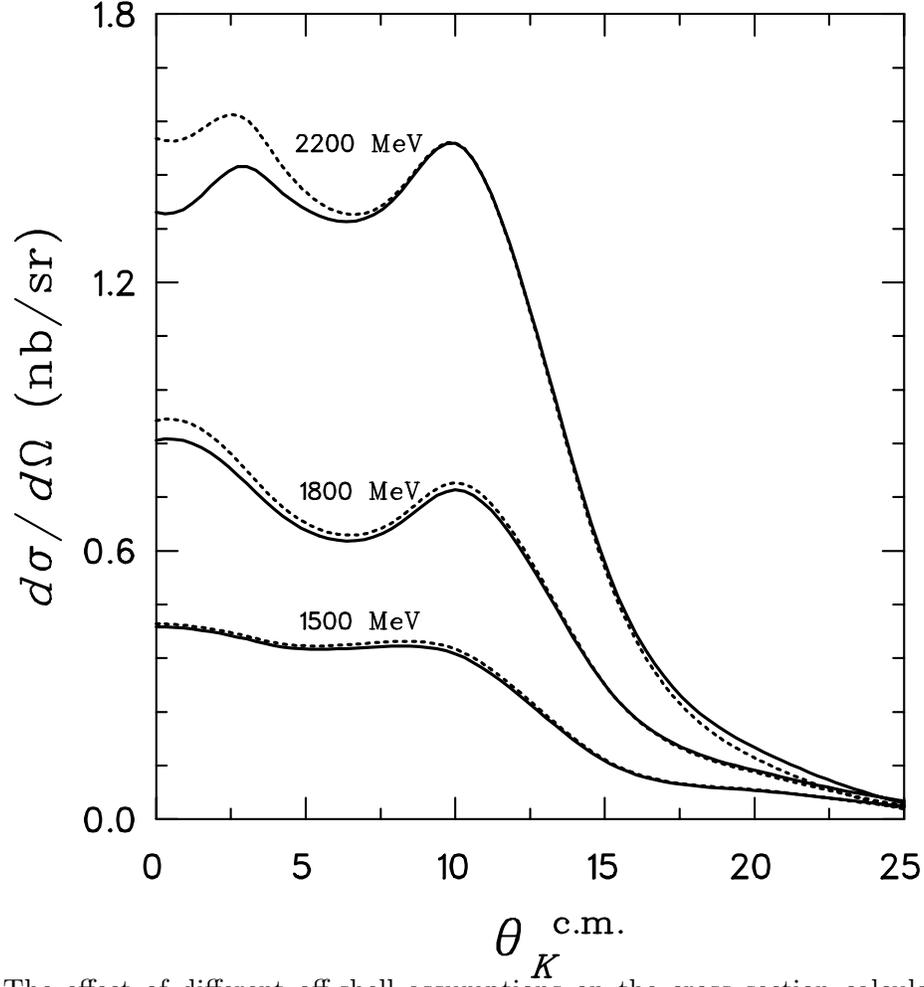,width=12cm}}
\caption{The effect of different off-shell assumptions on the cross section
 calculated at three different energies.
 The dotted curves have been calculated with the initial nucleon 
 on-shell $\left[ k_1^0 = (m_p^2+{\vec k}_1^2)^{1/2} \right]$, 
 the solid curves with the final hyperon  on-shell
 $\left[ k_{1'}^0 = (m_\Lambda^2+{\vec k}_{1'}^2)^{1/2} \right]$.}
\label{fvaros}
\end{figure}

Finally, we show the effect of different off-shell assumptions on the
cross section in Fig.~\ref{fvaros}. During the process, the nucleons 
in the initial and final states are clearly off-shell. However, the 
elementary amplitudes are in principle only valid for on-shell nucleons in the 
initial and final states. For this reason, we test the prescriptions 
given in Ref.~\cite{tiator2}, i.e. we assume that (1) the initial
nucleon is on-shell $\left( k_1^0 = [m_p^2+{\vec k}_1^2]^{1/2} \right)$,
the final hyperon is off-shell ($k_{1'}^0 = k_1^0 + |{\vec k}| - E_K$), 
and (2) the final hyperon is on-shell $\left( k_{1'}^0 = 
[m_{\Lambda}^2+{\vec k}_{1'}^2]^{1/2} \right)$, the initial nucleon is 
off-shell ($k_1^0 = k_{1'}^0 + E_K - |{\vec k}|$). Both assumptions are 
compared in Fig.~\ref{fvaros}, where we see that the difference is not too 
significant. The largest discrepancy of 10\% occurs at $k=2200$ MeV 
in the forward direction. The same behavior was found in the case 
of pion photoproduction, where the excitation energy is far below 
our energy of interest. 

Coulomb corrections, included as in Ref.~\cite{tia_th}, are found to have 
a negligible effect on our results. The inclusion of this effect 
decreases the cross section at forward angles by less than 4\%. This is 
in contrast to pion photoproduction, where the Gamow factor yields a 
significant reduction of the total cross section at threshold \cite{tia_th}.

\section{SUMMARY AND CONCLUSION}
\label{section_summary}
In this paper, we have presented the first cross section calculations 
for kaon photoproduction on $^3$He in the framework of the impulse
approximation. Apart from the non-relativistic reduction of the
amplitudes, we used the same method which has been successfully 
used to study pion photo- and electroproduction on $^3$He. 
The interesting feature offered by kaon production is the study 
of the hypertriton, the lightest and most loosely bound hypernucleus. 
In our study we used a $^3$He wave function from solutions of the 
Faddeev equations and a simple model for the hypertriton wave function. 
The predicted cross sections are small, about 3 nb/sr at forward directions.
Our results are compatible with an analysis of 
the hypertriton production through proton--deuteron collisions. We have 
also shown that the most promising kinematics for the corresponding 
experiment is at forward angles, where the momentum transfer 
reaches its minimum at high photon energies. 

In order to observe this process at Jefferson Lab, one may have to 
observe the hypertriton weak decay along with the detection of kaons. 
There are two modes of decay for the hypertriton, the mesonic channels 
$^3_{\Lambda}{\mathrm H} \rightarrow \pi + X$, and the non-mesonic
one $^3_{\Lambda}{\mathrm H} \rightarrow p+n+n$. A Monte Carlo study on
the kinematics of the electromagnetic production of the hypertriton 
\cite{schumacher2} shows that the mesonic mode would be difficult
to observe. Thus, only the non-mesonic decays could serve as a signal 
of hypertriton formation, leading to a very difficult experiment 
since only a tiny fraction would be taggable in this way \cite{schumacher2}.

From a theoretical point of view, it would also be interesting 
to investigate the production through a virtual photon, since the 
longitudinal component of the virtual photon would give additional 
information. In the case of pion electroproduction off $^3$He, it 
has been shown that the effects of Fermi motion and off-shell 
assumptions are larger than in photoproduction \cite{tiator2x}. 
As an example, the average momentum approximation can overestimate 
the transverse cross section for pion electroproduction by as much as 30\%.

Finally, we plan to study the quasi-free production of the lambda 
(i.e. the break-up process) $\gamma + {^3{\mathrm He}} \rightarrow
K^+ + \Lambda + NN$ in the future. This process is expected to be more
likely than the hypertriton production, because it does not require the 
formation of a bound state at high momentum transfer. Consequently, the 
corresponding cross sections should be significantly larger than
in the case of hypertriton formation. The $K^+$ quasi-free production
on $^3$He will be an important testing ground for $\Lambda NN$ continuum
3-body wave functions as well as $\Lambda NN$ 3-body force effects.

\acknowledgments

We are grateful to R. A. Schumacher for useful conversations and to S. S. 
Kamalov for his help with some of the approximations. We thank W. Gl\"ockle
for his help in normalizing the three-body wave functions and K. Miyagawa
for providing the hypertriton wave functions.
This work was supported by Deutscher Akademischer Austauschdienst, 
Deutsche Forschungsgemeinschaft (SFB 201), US Department of Energy 
grant no. DE-FG02-95-ER40907, and University Research for Graduate
Education (URGE) grant.

\newpage
\appendix

\section{THE NON-RELATIVISTIC OPERATOR}
\label{appnonrel}
The transition operator for the reaction $(e,e'K^+)$ is given by \cite{deo}
\begin{eqnarray}
{\cal M}_{\mathrm fi} & = & \bar u(p_{Y}) \sum_{i=1}^6 A_{i} M_{i} ~ u(p_{N})~ .
\end{eqnarray}
The amplitudes $A_{i}$ can be obtained from suitable Feynman 
diagrams for the
elementary reaction, while the gauge and Lorentz invariant matrices 
$M_{i}$ are given by \cite{terry}
\begin{eqnarray}
M_{1} &=& {\textstyle \frac{1}{2}} \gamma_{5} \left( \epsilon \!\!/ k 
\!\!\!/ - k \!\!\!/ \epsilon \!\!/ \right)~ ,\\
M_{2} &=& \gamma_{5}\left[ (2q_K-k) \cdot \epsilon P \cdot k - (2q_K-k) 
\cdot k P \cdot \epsilon \right]~ ,\\
M_{3} &=& \gamma_{5} \left( q_K\cdot k \epsilon \!\!/ - q_K\cdot \epsilon k
\!\!\!/ \right) ~ ,\\
M_{4} &=& i \epsilon_{\mu \nu \rho \sigma} \gamma^{\mu} q_K^{\nu}
\epsilon^{\rho} k^{\sigma}~ ,\\
M_{5} &=& \gamma_{5} \left( q_K\cdot \epsilon k^{2} - q_K\cdot k k \cdot 
\epsilon \right) ~ ,\\
M_{6} &=& \gamma_{5} \left( k \cdot \epsilon k \!\!\!/ - k^{2} \epsilon \!\!/
\right)~ .
\end{eqnarray}
The transition operator 
can be reduced into Pauli space in the case of free Dirac spinors, 
\begin{eqnarray}
\lefteqn{ \overline{u}({\vec p}_{Y})~\sum_{i=1}^{6}~ A_{i}M_{i}~u({\vec p}_{N})
 ~=~ \left(\frac{E_{N} + m_{N}}{2m_{N}} \right)^{\frac{1}{2}}
\left(\frac{E_{Y} + m_{Y}}{2m_{Y}} \right)^{\frac{1}{2}} \times }\nonumber\\
 & & \chi_{\mathrm f}^{\dagger}~\biggl[~{\cal{F}}_{1}
~{\vec{\sigma}} \cdot {\vec{\epsilon}}
+ {\cal{F}}_{2}~  {\vec{\sigma}} \cdot {\vec k}~ \epsilon_{0}
+ {\cal{F}}_{3}~ {\vec{\sigma}} \cdot {\vec k}~ 
{\vec k} \cdot  {\vec{\epsilon}} + {\cal{F}}_{4}~ 
{\vec{\sigma}} \cdot {\vec k}~ {\vec p}_N \cdot 
{\vec{\epsilon}} + {\cal{F}}_{5}~ {\vec{\sigma}}
\cdot {\vec k}~ {\vec p}_Y \cdot {\vec{\epsilon}}
\nonumber\\ 
&& +~ {\cal{F}}_{6}~ {\vec{\sigma}}\cdot {\vec p}_N~\epsilon_{0} 
+ {\cal{F}}_{7}~ {\vec{\sigma}} 
\cdot {\vec p}_N~ {\vec k} \cdot {\vec{\epsilon}}
+ {\cal{F}}_{8}~ {\vec{\sigma}} \cdot {\vec p}_N~ {\vec p}_N 
\cdot {\vec{\epsilon}} + {\cal{F}}_{9}~ 
{\vec{\sigma}} \cdot {\vec p}_N~ 
{\vec p}_Y \cdot  {\vec{\epsilon}}
\nonumber\\ 
 & & +~ {\cal{F}}_{10}~ {\vec{\sigma}} \cdot {\vec p}_Y~ 
\epsilon_{0} 
+ {\cal{F}}_{11}~ {\vec{\sigma}} \cdot {\vec p}_Y~ {\vec k} \cdot 
{\vec{\epsilon}} + {\cal{F}}_{12}~ {\vec{\sigma}}
 \cdot {\vec p}_Y~ {\vec p}_N \cdot {\vec{\epsilon}} + 
{\cal{F}}_{13}~ {\vec{\sigma}} \cdot {\vec p}_Y~ 
{\vec p}_Y \cdot  {\vec{\epsilon}}
\nonumber\\ 
 & & +~ {\cal{F}}_{14}~ {\vec{\sigma}} \cdot 
{\vec{\epsilon}}~ {\vec{\sigma}}
 \cdot {\vec k}~ {\vec{\sigma}} \cdot {\vec p}_N
 + {\cal{F}}_{15}~ {\vec{\sigma}} \cdot {\vec p}_Y~ 
{\vec{\sigma}} \cdot {\vec{\epsilon}}~ {\vec{\sigma}} \cdot {\vec k}
+ {\cal{F}}_{16}~ {\vec{\sigma}} \cdot {\vec p}_Y~ 
{\vec{\sigma}} \cdot {\vec{\epsilon}}~ {\vec{\sigma}} \cdot {\vec p}_N 
\nonumber\\ 
 & & +~ {\cal{F}}_{17}~ {\vec{\sigma}} \cdot {\vec p}_Y~ {\vec{\sigma}} \cdot 
{\vec k}~ {\vec{\sigma}} \cdot {\vec p}_N~ \epsilon_{0} 
+ {\cal{F}}_{18}~ {\vec{\sigma}} \cdot {\vec p}_Y~ 
{\vec{\sigma}} \cdot {\vec k}~ {\vec{\sigma}} 
\cdot {\vec p}_N~ {\vec k} \cdot {\vec{\epsilon}}
\nonumber\\ 
 & & +~ {\cal{F}}_{19}~ {\vec{\sigma}} \cdot {\vec p}_Y~ 
{\vec{\sigma}} \cdot {\vec k}~ {\vec{\sigma}} 
\cdot {\vec p}_N~ {\vec p}_N \cdot  {\vec{\epsilon}}
+ {\cal{F}}_{20}~ {\vec{\sigma}} \cdot {\vec p}_Y~ 
{\vec{\sigma}} \cdot {\vec k}~ {\vec{\sigma}} 
\cdot {\vec p}_N~ {\vec p}_Y \cdot  {\vec{\epsilon}}
~ \biggr] ~ \chi_{\mathrm i} ~,
\label{nro1}
\end{eqnarray}
where the individual amplitudes ${\cal F}_{i}$ are given by 
\begin{eqnarray}
{\cal{F}}_{1} & = & k_{0}A_{1} + k \cdot q_K A_{3} + \{2P \cdot k - 
k_{0}(m_{N} + m_{Y})\}A_{4} - k^{2}A_{6} ~,\\
{\cal{F}}_{2} & = & -A_{1} - (E_N+k_0-E_Y) A_{3} - (E_{N} + E_{Y} - m_{N} - 
m_{Y})A_{4} + k_{0}A_{6} ~, \\
{\cal{F}}_{3} & = & A_{3} - A_{6} ~, \\
{\cal{F}}_{4} & = & A_{3} + A_{4} ~, \\
{\cal{F}}_{5} & = & -A_{3} + A_{4} ~, \\
{\cal{F}}_{6} & = & \frac{1}{E_{N} + m_{N}}~\Bigl[~\{2P \cdot 
k(E_{N} - E_{Y})+({\textstyle \frac{1}{2}} k^{2} - k \cdot q_K)(E_{N} + E_{Y})
+ P \cdot kk_{0}\}A_{2} \nonumber\\
 && \hspace{2.2cm} + \{ k_0(E_N+k_0-E_Y) - k \cdot q_K \} A_{3} + 
\{k_{0}(E_{N} + E_{Y})  
\nonumber\\
 & & \hspace{2.2cm} - 2P \cdot k \}A_{4} - \{ k_0 k \cdot q_K - 
k^2(E_N+k_0-E_Y) \} A_{5} \nonumber\\
 & &  \hspace{2.2cm} + (k^2 - k_0^2) A_{6}~\Bigr] ~, \\
{\cal{F}}_{7} & = & \frac{1}{E_{N} + m_{N}}~\Bigl[~A_{1} - P \cdot k A_{2} 
- k_{0}A_{3} - (m_{N} + m_{Y})A_{4} - (k^{2} - k \cdot q_K)A_{5} \nonumber\\
 & &\hspace{2.2cm} + k_{0}A_{6}~\Bigr] ~, \\
{\cal{F}}_{8} & = & \frac{1}{E_{N} + m_{N}}~\Bigl[-(2P \cdot k + 
{\textstyle \frac{1}{2}} k^{2} - k \cdot q_K)A_{2} - k_{0}(A_{3} + A_{4}) - 
k^{2}A_{5}~\Bigr] ~, \\
{\cal{F}}_{9} & = & \frac{1}{E_{N}+m_{N}}~\Bigl[~(2P \cdot k - 
{\textstyle \frac{1}{2}} k^{2} + k \cdot q_K)A_{2} + k_{0}(A_{3} - A_{4}) + 
k^{2}A_{5}~\Bigr] ~, \\
{\cal{F}}_{10} & = & \frac{1}{E_{Y} + m_{Y}}~\Bigl[-\{2P \cdot k(E_{N} 
- E_{Y}) + ({\textstyle \frac{1}{2}} k^{2} - k \cdot q_K)(E_{N} + 
E_{Y}) + P \cdot k k_{0} \}A_{2}\nonumber\\
 & & \hspace{2.1cm}  + \{ k_0(E_N+k_0-E_Y) - 
k \cdot q_K \} A_{3} + \{k_{0}(E_{N} + E_{Y})  
\nonumber\\
 & & \hspace{2.1cm}  - 2P \cdot k \}A_{4} + \{ k_0 k \cdot q_K - 
k^2(E_N+k_0-E_Y) \} A_{5} \nonumber\\
 & & \hspace{2.2cm} + (k^2-k_0^2) A_{6}~\Bigr] ~, \\
{\cal{F}}_{11} & = & \frac{1}{E_{Y} + m_{Y}}~\Bigl[-A_{1} + P \cdot kA_{2} 
- k_{0}A_{3} + (m_{N} + m_{Y})A_{4} + (k^{2} - k \cdot q_K)A_{5} 
\nonumber\\
 & & \hspace{2.1cm} + k_{0}A_{6}~\Bigr] ~, \\
{\cal{F}}_{12} & = & \frac{1}{E_{Y} + m_{Y}}~\Bigl[~(2P \cdot k + 
{\textstyle \frac{1}{2}} k^{2} - k \cdot q_K)A_{2} - k_{0}(A_{3} + A_{4}) + 
k^{2}A_{5}~\Bigr] ~, \\
{\cal{F}}_{13} & = & \frac{1}{E_{Y} + m_{Y}}~\Bigl[-(2P \cdot k 
+ k \cdot q_K - {\textstyle \frac{1}{2}} k^{2})A_{2} + k_{0}(A_{3} - A_{4}) 
- k^{2}A_{5}~\Bigr] ~, \\
{\cal{F}}_{14} & = & \frac{1}{E_{N} + m_{N}}~\Bigl[-A_{1}
 + (m_{N} + m_{Y})A_{4}~\Bigr] ~, \\
{\cal{F}}_{15} & = & \frac{1}{E_{Y} + m_{Y}}~\Bigl[~A_{1}
 - (m_{N} + m_{Y})A_{4}~\Bigr] ~, \\
{\cal{F}}_{16} & = & \frac{1}{(E_{N} + m_{N})(E_{Y} + m_{Y})} 
~\Bigl[-k_{0}A_{1} + k \cdot q_K A_{3} + \{2P \cdot k + k_{0}(m_{N} + 
m_{Y})\}A_{4} \nonumber\\
 & & \hspace{4.2cm} - k^{2}A_{6}~\Bigr] ~, \\
{\cal{F}}_{17} & = & \frac{1}{(E_{N} + m_{N})(E_{Y} + m_{Y})}~\Bigl[~A_{1} 
- (E_N+k_0-E_Y) A_{3} - (E_{N} + E_{Y} \nonumber\\ 
 & & \hspace{4.2cm} + m_{N} + m_{Y})A_{4} + k_{0}A_{6}~\Bigr] ~, \\
{\cal{F}}_{18} & = & \frac{1}{(E_{N} + m_{N})(E_{Y} + m_{Y})} 
~\Bigl[~A_{3} - A_{6}~\Bigr] ~, \\
{\cal{F}}_{19} & = & \frac{1}{(E_{N} + m_{N})(E_{Y} + m_{Y})} 
~\Bigl[~A_{3} + A_{4}~\Bigr] ~, \\
{\cal{F}}_{20} & = & \frac{1}{(E_{N} + m_{N})(E_{Y} + m_{Y})} 
~\Bigl[-A_{3} + A_{4}~\Bigr] ~.
\end{eqnarray}

The spin-independent and spin-dependent amplitudes of 
Eq.~(\ref{lk1}) and (\ref{lk2}) can be derived from Eq.~(\ref{nro1})
by making use of the relation ${\vec{\sigma}} \cdot 
{\vec a}~{\vec{\sigma}} \cdot {\vec b} ~=~ {\vec a} \cdot 
{\vec b} + i~ {\vec{\sigma}} \cdot {\vec a} \times {\vec b}$,
yielding 
\begin{eqnarray}
L & = & N~ \Bigl\{ -\left( {\cal{F}}_{14}+{\cal{F}}_{15}-{\cal{F}}_{16}
\right)~ {\vec p}_N \cdot ({\vec k} \times {\vec{\epsilon}}\, ) 
+{\cal{F}}_{15}~{\vec q}_K\cdot ({\vec k} \times {\vec{\epsilon}\, })
\nonumber\\ && \hspace{7mm} - {\cal{F}}_{16}~{\vec p}_N \cdot 
({\vec q}_K \times {\vec{\epsilon}\, })
-\left[ {\cal{F}}_{17}~\epsilon_0 + \left({\cal{F}}_{18}+{\cal{F}}_{20}
\right)~ {\vec k} \cdot {\vec{\epsilon}} \right.
\nonumber\\ 
&& \hspace{7mm} \left.
 +\left( {\cal{F}}_{19}+{\cal{F}}_{20} \right)~ {\vec p}_N \cdot 
{\vec{\epsilon}} - {\cal{F}}_{20}~ {\vec q}_K \cdot {\vec k}
\right]~ {\vec p}_N \cdot ({\vec q}_K \times {\vec{\epsilon}\, })
\Bigr\} ~,\\
\nonumber\\
  {\vec K} &=& -N \left( T_1~{\vec{\epsilon}}+T_2~{\vec k}+
T_3~{\vec p}_N+  T_4~{\vec q}_K \right) ~,
\end{eqnarray}
with
\begin{eqnarray}
N &=& \left(\frac{E_{N} + m_{N}}{2m_{N}} \right)^{\frac{1}{2}}
\left(\frac{E_{Y} + m_{Y}}{2m_{Y}} \right)^{\frac{1}{2}} ~ ,
\end{eqnarray}
and
\begin{eqnarray}
  T_1 &=& {\cal{F}}_{1}+({\cal{F}}_{14}-{\cal{F}}_{15}-{\cal{F}}_{16})~
          {\vec p}_N \cdot {\vec k}+ {\cal{F}}_{15}~ ({\vec q}_K\cdot{\vec k}
          - {\vec k}^2) \nonumber\\
      && +{\cal{F}}_{16}~({\vec p}_N\cdot{\vec q}_K - {\vec p}_N^{\, 2}) ~,\\
      \nonumber\\
  T_2 &=& [{\cal{F}}_{2}+{\cal{F}}_{10}+({\vec p}_N\cdot{\vec q}- 
     {\vec p}_N^{\, 2})~{\cal{F}}_{17}]~\epsilon_0+[{\cal{F}}_{3}+{\cal{F}}_{5}
      +{\cal{F}}_{11}+{\cal{F}}_{13}+2{\cal{F}}_{15}\nonumber\\
      && +({\vec p}_N \cdot {\vec q}-{\vec p}_N^{\, 2})~({\cal{F}}_{18}+
      {\cal{F}}_{20})]~{\vec k}\cdot{\vec{\epsilon}} 
      +[{\cal{F}}_{4}+
      {\cal{F}}_{5}+{\cal{F}}_{12}+{\cal{F}}_{13}-{\cal{F}}_{14}
      \nonumber\\
      && +{\cal{F}}_{15}+{\cal{F}}_{16}+({\vec p}_N \cdot {\vec q}-
      {\vec p}_N^{\, 2})~({\cal{F}}_{19}+
      {\cal{F}}_{20})]~{\vec p}_N\cdot{\vec{\epsilon}}
      -[{\cal{F}}_{5}+{\cal{F}}_{13}+{\cal{F}}_{15}\nonumber\\
      && +({\vec p}_N\cdot{\vec q}-{\vec p}_N^{\, 2})~
      {\cal{F}}_{20}]~{\vec q}_K\cdot{\vec{\epsilon}} ~,\\
      \nonumber\\
  T_3 &=& [{\cal{F}}_{6}+{\cal{F}}_{10}+(2{\vec p}_N\cdot{\vec k}+ 
      {\vec k}^2-{\vec q}_K\cdot{\vec k})~{\cal{F}}_{17}]~\epsilon_0+
      [{\cal{F}}_{7}+{\cal{F}}_{9}+{\cal{F}}_{11}+{\cal{F}}_{13}\nonumber\\
      &&+{\cal{F}}_{14}+{\cal{F}}_{15}+{\cal{F}}_{16} +(2{\vec p}_N \cdot 
      {\vec k}+{\vec k}^2-{\vec q}_K\cdot{\vec k})~ ({\cal{F}}_{18}+
      {\cal{F}}_{20})]~{\vec k}\cdot{\vec{\epsilon}}
      + [{\cal{F}}_{8} \nonumber\\
      &&+{\cal{F}}_{9}+{\cal{F}}_{12}+{\cal{F}}_{13}+2{\cal{F}}_{16}
      +(2{\vec p}_N \cdot {\vec k}+{\vec k}^2-{\vec q}_K\cdot{\vec k})~
     ({\cal{F}}_{19}+{\cal{F}}_{20})]~{\vec p}_N\cdot
      {\vec{\epsilon}}\nonumber\\
      && -[{\cal{F}}_{9}+{\cal{F}}_{13}+{\cal{F}}_{16}
      +(2{\vec p}_N \cdot {\vec k}+{\vec k}^2-{\vec q}_K\cdot{\vec k})~
      {\cal{F}}_{20}]~{\vec q}_K\cdot{\vec{\epsilon}} ~,\\
      \nonumber\\
  T_4 &=& -({\cal{F}}_{10}+{\cal{F}}_{17}~{\vec p}_N \cdot {\vec k})~
      \epsilon_0-[{\cal{F}}_{11}+{\cal{F}}_{13}+{\cal{F}}_{15}+
      ({\cal{F}}_{18}+{\cal{F}}_{20})~{\vec p}_N \cdot {\vec k}]~
      {\vec k}\cdot{\vec{\epsilon}}\nonumber\\
      &&+[{\cal{F}}_{12}+{\cal{F}}_{13}+{\cal{F}}_{16}+
      ({\cal{F}}_{19}+{\cal{F}}_{20})~{\vec p}_N \cdot {\vec k}]~
      {\vec p}_N\cdot{\vec{\epsilon}} \nonumber\\
      &&+({\cal{F}}_{13}+{\cal{F}}_{20}~
      {\vec p}_N\cdot{\vec k})~{\vec q}_K\cdot{\vec{\epsilon}} ~.
\end{eqnarray}
Finally, after neglecting the small terms ${\cal{F}}_{16}$ - 
${\cal{F}}_{20}$, and also dropping the terms containing $k^2$, ${\vec k} 
\cdot {\vec{\epsilon}}$, and $\epsilon_0$, 
we obtain Eq.~(\ref{lk1}) and (\ref{lk2}). 
Gauge invariance can be checked by observing
\begin{eqnarray}
\label{testgi1}
  L({\epsilon} \rightarrow k) &=& 0 ~,\\
  {\vec K}({\epsilon} \rightarrow k) &=& 0 ~.
\label{testgi2}
\end{eqnarray}
We note that Eq.~(\ref{testgi1}) and (\ref{testgi2}) are still satisfied
after the omission of ${\cal{F}}_{16}$ - ${\cal{F}}_{20}$.

\begin{table}[htbp]
\begin{center}
\caption{Quantum numbers and probabilities of the $^3{\mathrm He}$ and 
$^3_{\Lambda}{\mathrm H}$ wave functions.}
\label{phewf}
\begin{tabular}{rrrrrrrrrrccccccc}
$\alpha$ &&$L$ && $l$ && ${\cal L}$ && $S$ && ${\cal S}$ && $T$ &&
$P(^3{\mathrm He})(\%)$&$P(^3_{\Lambda}{\mathrm H})(\%)$&\\
[1.5ex]
\tableline
~~ 1& ~ &0& ~ &0& ~ &0& ~ &1& ~ &$1/2$& ~ &0&&44.31&94.23&\\
2&&0&&0&&0&&0&&$1/2$&&1&&43.70&-&\\
3&&2&&2&&0&&1&&$1/2$&&0&&\enspace 0.47&-&\\
4&&2&&2&&0&&0&&$1/2$&&1&&\enspace 0.90&-&\\
5&&1&&1&&0&&0&&$1/2$&&0&&\enspace 0.41&-&\\
6&&1&&1&&0&&1&&$1/2$&&1&&\enspace 0.41&-&\\
7&&2&&0&&2&&1&&$3/2$&&0&&\enspace 3.06&\enspace 5.77&\\
8&&0&&2&&2&&1&&$3/2$&&0&&\enspace 1.00&-&\\
9&&1&&1&&2&&1&&$3/2$&&1&&\enspace 2.40&-&\\
10&&3&&1&&2&&1&&$3/2$&&1&&\enspace 0.39&-&\\
11&&1&&3&&2&&1&&$3/2$&&1&&\enspace 1.06&-&\\
[2.1ex]
\end{tabular}
\end{center}
\end{table}

\begin{table}[htbp]
\begin{center}
\caption{Some characteristics of the triton 
($^{3}$H), hypertriton ($^{3}_{\Lambda}$H), and deuteron ($d$).}
\label{tb}
\begin{tabular}{lccc}
 & $^{3}$H~\protect\cite{tiley} & $^{3}_{\Lambda}$H & $d$ \\ 
[1.5ex]
 \tableline
bind. energy (MeV) & 8.481855 & 0.13$\pm$0.05~\protect\cite{juric1}& 2.224575\\
charge ($e^{+}$) & 1  & 1 & 1\\
spin ($J^{\pi}$) & ~ $\frac{1}{2}^{+}$ & ~ $\frac{1}{2}^{+}$  & ~ $1^{+}$ \\
isospin ($I$) & $\frac{1}{2}$   & 0  & 0 \\
magn. moment ($\mu_{N}$) & 2.97896(1) &0.78~\cite{dover}& 0.857406
\cite{honzawa}\\
half life & 12.32 years & (2.64 - 0.95) $10^{-10}$ s \cite{oldi} &stable\\
mass (MeV) & 2808.94 & 2991.11 & 1875.61\\
[2.1ex]
\end{tabular}
\end{center}
\end{table}

\begin{table}[htbp]
\begin{center}
\caption{Tensor operators $\left[ {\bf Y}^{(l)}(\hat{{\vec q}\, }) \otimes 
{\bf K}^{(n)} \right]^{(\Lambda )}_{m_{\Lambda}}$.}
\label{tensor}
\begin{tabular}{ccccccclc}
~~~~&$l$&~~~&$n$&~~~&$\Lambda$&~\hspace{2cm}~&$\left[ {\bf Y}^{(l)}
(\hat{{\vec q}\, }) 
\otimes {\bf K}^{(n)} \right]^{(\Lambda )}_{m_{\Lambda}}$&~~~~\\
[1.5ex]
\tableline\\
&0&&0&&0&&$\displaystyle \frac{1}{\sqrt{4\pi}}~L$&\\\\
&0&&1&&1&&$\displaystyle \frac{1}{\sqrt{4\pi}}~{\vec K}$&\\\\
&1&&0&&1&&$\displaystyle \sqrt{\frac{3}{4\pi}}~\hat{{\vec q}\, }L$&\\\\
&1&&1&&0&&$\displaystyle -\frac{1}{\sqrt{4\pi}}~\hat{{\vec q}\, } \cdot 
{\vec K}$&\\\\
&1&&1&&1&&$\displaystyle \sqrt{\frac{3}{8\pi}}~i\hat{{\vec q}\, } \times 
{\vec K}$&\\\\
&2&&1&&1&&$\displaystyle \sqrt{\frac{1}{8\pi}}~({\vec K}-3\hat{{\vec q}\, } 
\cdot {\vec K}~\hat{{\vec q}\, })$&\\
[2.1ex]
\end{tabular}
\end{center}
\end{table}

\begin{table}[htb]
\begin{center}
\caption{Coupling constants for different elementary models. Set I obtained
         by fitting to $K\Lambda$ photoproduction data \protect\cite{adel2}, 
         set II fits to both photo- and electroproduction data
         \protect\cite{adel1}, set III elementary
         model of Ref.~\protect\cite{williams} describing both  
         photo- and electroproduction data, set IV 
         model fitting all existing kaon
         photo- and electroproduction data \protect\cite{terry} by using 
         hadronic form factors. In set IV only the coupling constants 
         for the $K\Lambda$ channel are shown. $\Lambda_{\mathrm c}$ indicates
         the cut-off parameter used for the hadronic form factors.}
\label{newcc}
\begin{tabular}{lrrrr}
 Coupling Constants  & I  & II & III & IV \\ 
[1.5ex]
\tableline
$g_{K\Lambda N}/\sqrt{4\pi}$ &$-4.17$ &3.15 &$-2.38$ &$-3.09$\\
$g_{K\Sigma N}/\sqrt{4\pi}$ &1.18 &1.68 &0.23 & 1.22 \\
$G_{V}^{K^{*}}(892)/4\pi$ &$-0.43$ &0.03 &$-0.16$ & $-0.19$ \\
$G_{T}^{K^{*}}(892)/4\pi$ &0.20 &$-0.19$ &0.08 &$-0.12$\\
$G_{N1}(1440)/\sqrt{4\pi}$&$-1.79$ &$-1.11$ &- & -\\
$G_{N4}(1650)/\sqrt{4\pi}$& - &0.10 &$-0.06$ & $-0.06$\\
$G_{N6}(1710)/\sqrt{4\pi}$ &- &- &$-0.09$ & $-0.07$ \\ 
$G_{V}^{K_1}(1270)/4\pi$ &$-0.10$ &0.13 &0.02 & - \\
$G_{T}^{K_1}(1270)/4\pi$ &$-1.21$ &0.06 &0.17 &-\\
$G_{Y1}(1405)/\sqrt{4\pi}$&- & - &$-0.10$ & -\\
$G_{Y3}(1670)/\sqrt{4\pi}$&$-4.71$ &0.70 &- & -\\
$\Lambda_{\mathrm c}$ & - & - & - & 0.85\\
[2.1ex]
\end{tabular}
\end{center}
\end{table}

\vspace{2cm}

\vspace{15mm}

\end{document}